\documentclass[twocolumn,trackchanges]{aastex61}

\usepackage{amsmath,amstext,float}
\usepackage[T1]{fontenc}
\usepackage{apjfonts} 
\usepackage[figure,figure*]{hypcap}
\usepackage{CJKutf8}

\newcommand{\HII}{\ion{H}{2}}

\newcommand{\lzifu} {{\scshape lzifu}}
\newcommand{\ppxf} {{\scshape ppxf}}
\newcommand{\idl} {{\scshape idl}}
\newcommand{\izi} {{\scshape izi}}
\newcommand{\mappingsiv} {{\scshape mappings~iv}}
\newcommand{\mappingsv} {{\scshape mappings~v}}
\newcommand{\hiiphot} {{\scshape hiiphot}}

\shorttitle{Azimuthal variations of abundance in NGC\,1365}
\shortauthors{Ho et al.}

\begin{document}
\begin{CJK*}{UTF8}{gbsn}

\title{The chemical evolution carousel of spiral galaxies: azimuthal variations of oxygen abundance in NGC\,1365}

\author[0000-0002-0757-9559]{I-Ting Ho (何宜庭)}
\affiliation{Max Planck Institute for Astronomy, K\"{o}nigstuhl 17, 69117 Heidelberg, Germany}
\author{Mark Seibert}
\affiliation{Observatories of the Carnegie Institution for Science, 813 Santa Barbara Street, Pasadena, CA 91101, USA}
\author{Sharon E. Meidt}
\affiliation{Max Planck Institute for Astronomy, K\"{o}nigstuhl 17, 69117 Heidelberg, Germany}
\author{Rolf-Peter Kudritzki}
\affiliation{Institute for Astronomy, University of Hawaii, 2680 Woodlawn Drive, Honolulu, HI 96822, USA}
\affiliation{University Observatory Munich, Scheinerstr. 1, D-81679 Munich, Germany}
\author{Chiaki Kobayashi}
\affiliation{Centre for Astrophysics Research, School of Physics, Astronomy and Mathematics, University of Hertfordshire, College Lane, Hatfield AL10 9AB, UK}
\author{Brent A. Groves}
\affiliation{Research School of Astronomy and Astrophysics, Australian National University, Canberra 2600, ACT, Australia}
\author{Lisa J. Kewley}
\affiliation{Research School of Astronomy and Astrophysics, Australian National University, Canberra 2600, ACT, Australia}
\author{Barry F. Madore}
\affiliation{Observatories of the Carnegie Institution for Science, 813 Santa Barbara Street, Pasadena, CA 91101, USA}
\author{Jeffrey A. Rich}
\affiliation{Observatories of the Carnegie Institution for Science, 813 Santa Barbara Street, Pasadena, CA 91101, USA}
\author{Eva Schinnerer}
\affiliation{Max Planck Institute for Astronomy, K\"{o}nigstuhl 17, 69117 Heidelberg, Germany}
\author{Joshua D'Agostino}
\affiliation{Research School of Astronomy and Astrophysics, Australian National University, Canberra 2600, ACT, Australia}
\author{Henry Poetrodjojo}
\affiliation{Research School of Astronomy and Astrophysics, Australian National University, Canberra 2600, ACT, Australia}

\begin{abstract}
The spatial distribution of oxygen in the interstellar medium of galaxies is the key to understanding how efficiently metals that are synthesized in massive stars can be redistributed across a galaxy. We present here a case study in the nearby spiral galaxy NGC\,1365 using 3D optical data obtained in the TYPHOON Program. We find systematic azimuthal variations of the \HII\ region oxygen abundance imprinted on a negative radial gradient. The 0.2~dex azimuthal variations occur over a wide radial range of 0.3 to 0.7~$\rm R_{25}$ and peak at the two spiral arms in NGC\,1365. We show that the azimuthal variations can be explained by two physical processes: gas undergoes localized, sub-kpc scale self-enrichment when orbiting in the inter-arm region, and experiences efficient, kpc scale mixing-induced dilution when spiral density waves pass through. We construct a simple chemical evolution model to quantitatively test this picture and find that our toy model can reproduce the observations. This result suggests that the observed abundance variations in NGC\,1365 are a snapshot of the dynamical local enrichment of oxygen modulated by spiral-driven, periodic mixing and dilution.
\end{abstract}



\let\standardclearpage\clearpage
\let\clearpage\relax
\section{Introduction}
\let\clearpage\standardclearpage
The oxygen abundance of the interstellar medium (ISM) is one of the key physical properties for understanding galaxy evolution. The ISM oxygen abundance and distribution are built up by various physical processes throughout a galaxy's evolutionary history. Star formation, nucleosynthesis, stellar winds, outflows, and accretion can all change the ISM oxygen abundance. Although these physical processes are often coupled in complicated ways, chemical evolution models have been able to place quantitative constraints on the key processes driving galaxy evolution \citep[e.g.][]{Kobayashi:2007fk,Kobayashi:2011bv,Lilly:2013qf,Dave:2011qy,Torrey:2012ai,Zahid:2014uq,Taylor:2016aa}. 

\begin{table*}[t]
 \caption{NGC\,1365}
 \label{tbl1}
\centering
\begin{tabular}{lccc}
  \hline
  Property & Value & Note & Reference\\
  \hline
  Distance$^a$ & 18.1~Mpc & {\it HST} tip of red giant branch   & \citet{Jang:2017kq}\\
  Hubble type &  SB(s)b & -- & RC3 \citep{de-Vaucouleurs:1991sf} \\
  $\log M_*$ & 10.970 & Scaled from S$^4$G to the assumed distance & S$^4$G \citep{Munoz-Mateos:2013aa}\\
$b/a$ & 0.82 &  Axis-ratio of J+H+K image at $3\sigma$ isophote & 2MASS LGA \citep{Jarrett:2003ty}\\
  Inclination & $35.7^\circ$ & $\cos^2 {\rm inc.} = \frac{(b/a)^2 - q_0^2}{1 - q_0^2}$, assuming $q_0=0.2$ & \citet{Hubble:1926uq} \\
  P.A. & $49.5^\circ$ &  Position angle of J+H+K  isophote (east of north) & 2MASS LGA \citep{Jarrett:2003ty} \\
  $\rm R_{25}$ & 5.61\arcmin & Mean radius of 25 mag~arcsec$^{-2}$ isophote (B-band) & RC3 \citep{de-Vaucouleurs:1991sf}\\
  \hline
\multicolumn{4}{l}{$^a$ $1\rm \arcsec = 87.8~pc;\ 1\arcmin = 4.9~kpc.$}\\
 \end{tabular}
\end{table*}

Empirical relationships between the oxygen abundance of star-forming galaxies and their physical properties have been discovered on different physical scales in galaxies. The global oxygen abundance correlates strongly with the galaxy stellar mass, known as the ``mass-metallicity relation'' \citep{Lequeux:1979lr,Tremonti:2004fk}. The mass-metallicity relation has been measured out to $z>2$ and appears to evolve over cosmic time \citep[e.g.][]{Zahid:2013kx,Yuan2013aa,Sanders:2015uo}. On spatially-resolved scales, the radial distributions of the oxygen abundance in non-interacting galaxies can be characterized by negative abundance gradients \citep{Searle:1971kx,Zaritsky:1994lr}. This appears to reflect the inside-out formation history of the galactic disks. The slopes of this radial abundance gradient are similar for all non-interacting galaxies when normalized to their galaxy scale-length (e.g. effective or iso-photal radii; \citealt{Sanchez:2014fk,Ho:2015hl,Sanchez-Menguiano:2016vl}). Chemical evolution models suggest that the stellar and gas masses, and the ratio of the two are the key quantities driving the observed time evolution of the mass-metallicity relation and the presence of the ``common'' oxygen abundance gradient \citep{Zahid:2014uq,Kudritzki:2015fp,Ho:2015hl,Ascasibar:2015la}.

The fundamental processes governing ISM chemical enrichment operate on different scales. Oxygen is first produced mainly by high-mass stars ($>8\rm M_\odot$), then blown to parsec scales through winds or supernovae, and subsequently dispersed to kilo-parsec scales by time- and scale-dependent mixing processes. For example, turbulent diffusion, Rayleigh-Taylor and Kelvin-Helmhotz instabilities, and super-shell expansions of supernova remnants mix the ISM on sub-kpc scales (1 to 100~Myr). Turbulent transport in differentially rotating disks and bar can induce radial flows that further homogenize the ISM on kilo-parsec scales ($\lesssim 1$~Gyr; see \citealt{Roy:1995ys}). 

\begin{figure*}[!ht]
\centering
\includegraphics[width=\textwidth]{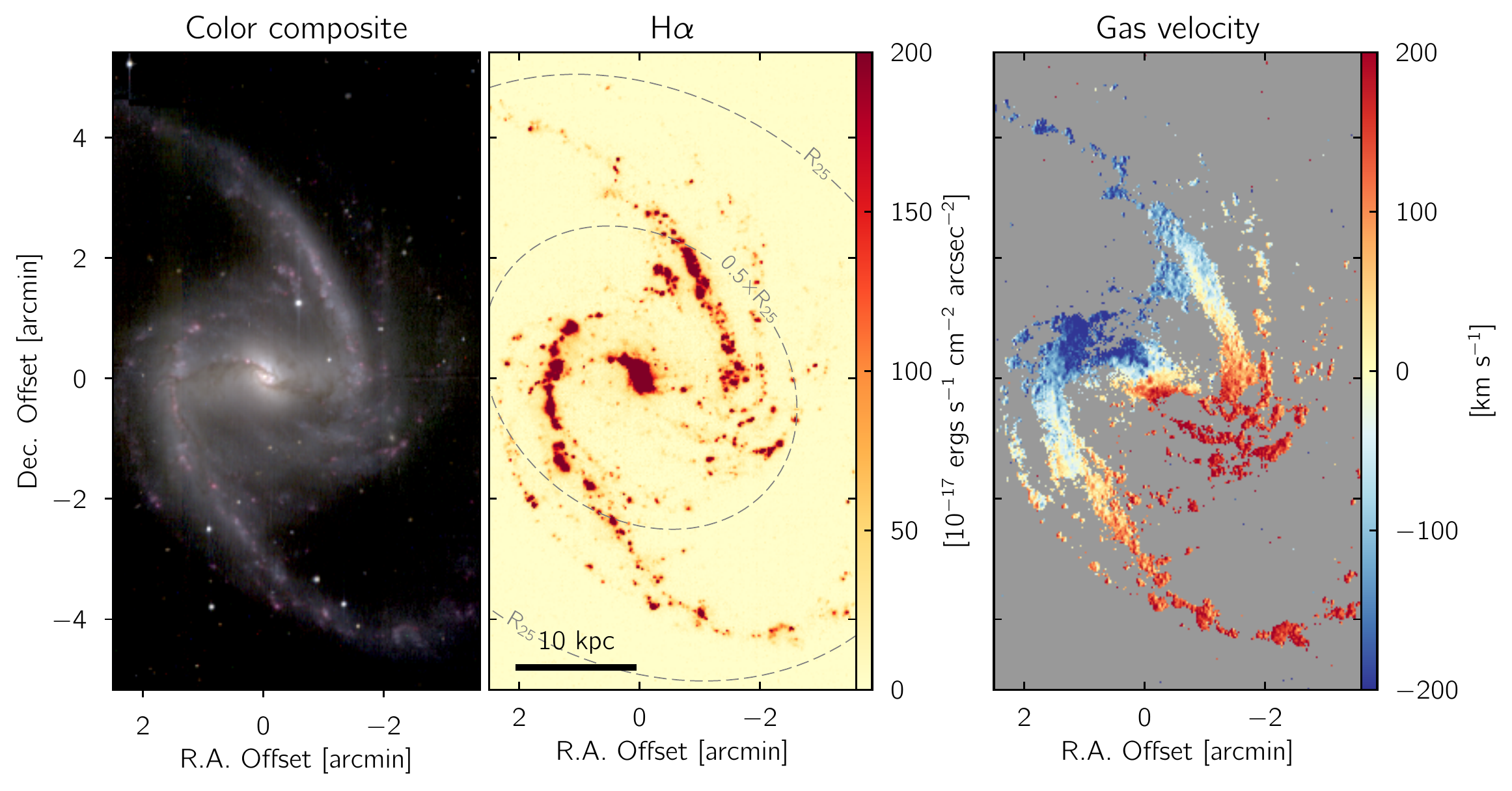}
\caption{
{\it Left}: {\it BVR} color composite image of NGC\,1365 reconstructed from the data cube. 
{\it Middle}: H$\alpha$ surface brightness distribution from fitting the emission line in the data cube. The two dashed ellipses correspond to $0.5\times$ and $1\times\rm R_{25}$ in the frame of the galactic disk. 
{\it Right}: Velocity field derived from fitting emission lines. }\label{fig1} 
\end{figure*}

At any instance, the spatial distribution of the ISM abundance represents a snapshot in time of the production history and mixing processes. To the zeroth order, the spatial distribution of oxygen abundance can be characterized by a simple gradient. Deviations from the simple gradient in the azimuthal direction could inform us on how oxygen produced by massive stars is mixed with their neighboring ISM when the gas and stars orbit in the galactic potential. The time-scale for differential rotation to chemically homogenize the ISM in the azimuthal direction ($\lesssim 1$~Gyr) is longer than both the oxygen production time-scale (OB star lifetime; $<10$~Myr) and the time-scales for mixing on sub-kpc scales (10 to 100~Myr; \citealt{Roy:1995ys}). Therefore one expects to observe some degree of azimuthal inhomogeneity of oxygen abundance.

Probing azimuthal inhomogeneities of chemical abundances has been attempted in the Milky Way and external galaxies. In the Milky Way, \citet{Balser:2011uo} measured \HII\ region oxygen abundances and found that the slopes of the oxygen abundance gradients differ by a factor of two in their three Galactic azimuth angle bins. Significant local iron abundance inhomogeneities have also been observed with Galactic Cepheids \citep{Pedicelli:2009qq,Genovali2014}. In nearby galaxies, spectroscopic studies of \HII\ region oxygen abundance have found little evidence for azimuthal variations, i.e.~the intrinsic scatter around the best-fit linear oxygen abundance gradients is consistent with zero  \citep{Martin:1996nr,Cedres:2002vl,Bresolin:2009xy,Bresolin:2011rt}. In a few cases, tentative evidence for azimuthal variations is reported \citep{Kennicutt:1996yf,Rosolowsky:2008qf,Li:2013nx,Kudritzki:2016zr}. 

Measuring and characterizing the azimuthal variations of the ISM oxygen abundance is challenging because there are large systematic uncertainties (relative to the intrinsic azimuthal variation) associated with determining ISM abundances, especially when only strong emission lines are measured \citep{Kewley:2008qy}. Furthermore, long-slit and multi-object spectroscopy studies usually pre-select bright \HII\ regions from narrow-band images and provide a biased view of azimuthal variations. The limited disk areal coverage of many surveys have prevented a complete view of the oxygen abundance distributions.

With the complete spatial coverage delivered by integral field spectroscopy (IFS), recent studies have shown evidence that the oxygen abundance varies azimuthally, and there appears to exist a spatial correlation between the azimuthal variations and spiral structures \citep[][see also \citealt{Cedres:2012ad} and \citealt{Croxall:2016tg}]{Sanchez-Menguiano:2016zr,Vogt:2017aa}. Using Multi Unit Spectroscopic Explorer (MUSE) on the Very Large Telescope (VLT), \citet{Sanchez-Menguiano:2016zr} show that the \HII\ region oxygen abundances are higher at the trailing edges and lower at the leading edges of the spiral arms of NGC\,6754. The H$\alpha$ velocity field is consistent with radially outward streaming motion at the trailing edges of the spiral arms and radially inward streaming motion at the leading edges. \citet{Sanchez-Menguiano:2016zr} conclude that it is the radial flows on both sides of the spiral arms that drive the azimuthal variations of oxygen abundance. \citet{Vogt:2017aa} present MUSE data of the compact group galaxy HCG~91c and show that,  between about 0.5 to $0.8R_{25}$, the oxygen abundance distribution correlates with the spiral arms. \citet{Kreckel:2016aa} find that in NGC\,628 the \HII\ regions in the inter-arm regions have oxygen abundances similar to those on the spiral arms. However, their MUSE data only cover the central region of NGC\,628 ($\lesssim 0.5R_{25}$).

The keys to detecting and correctly characterizing the azimuthal variations of the oxygen abundance in galaxies are complete spatial and spectral coverage. In this paper, we measure the azimuthal variations of oxygen abundance in the nearby spiral galaxy NGC\,1365 using integral field spectroscopy data obtained in the TYPHOON Program. We will show that NGC\,1365 presents coherent variations of oxygen abundance which spatially correlate with the spiral structures. Although this is not the first time azimuthal variations of oxygen abundance have been detected, NGC\,1365 presents, to the best of our knowledge, the largest and most pronounced azimuthal variations known to-date. We construct a simple chemical evolution model to understand the underlying physical processes governing the variations. We demonstrate that the variations can be explained by a combination of localized, star formation driven self-enrichment and large-scale mixing-driven dilution due to the passing of spiral density waves.

The paper is structured as follows. Section~2 describes our observations and data reduction. In Section~3, we introduce how we measure emission line fluxes and derive oxygen abundances. Section~4 presents the key results of the paper, including the pronounced azimuthal variations of oxygen abundance. In Section~5, we discuss what causes the observed azimuthal variations and propose a simple chemical model that can quantitatively explain our observations. Throughout the paper, we adopt the physical properties of NGC\,1365 listed in Table~\ref{tbl1}. 

\begin{figure*}[!ht]
\centering
\includegraphics[width=\textwidth]{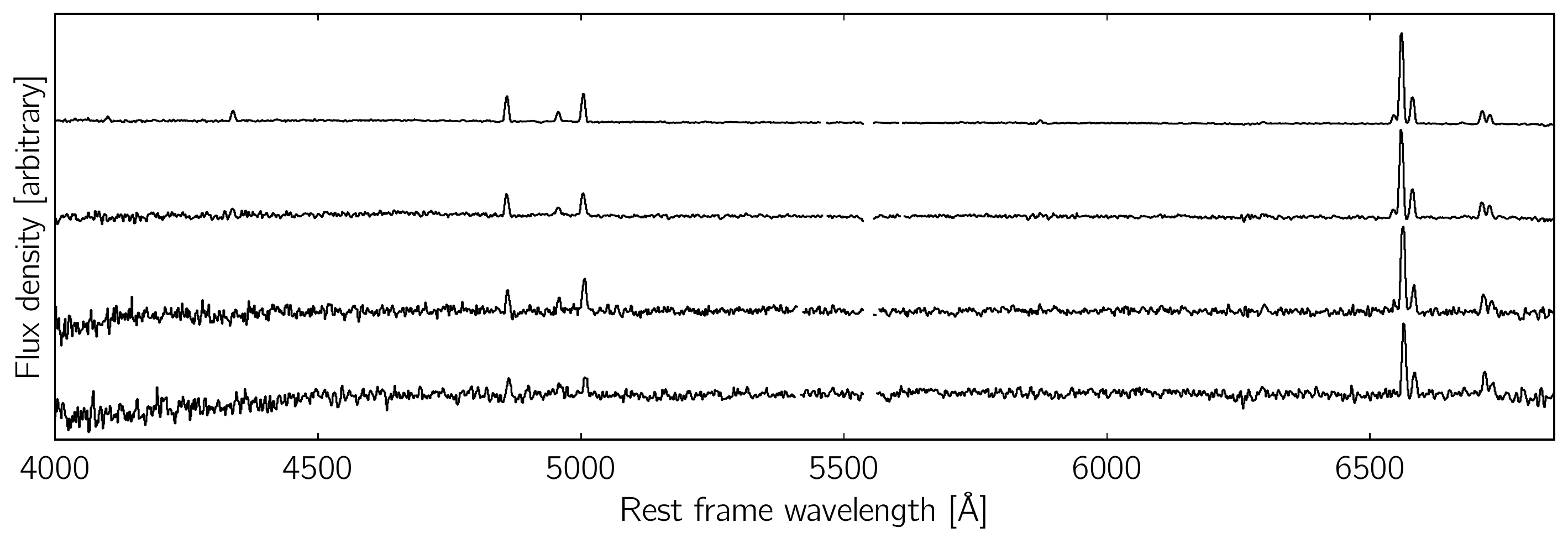}
\caption{Examples of the \HII\ region spectra. The four scaled spectra demonstrate our data quality from the first to the last S/N(H$\alpha$) quartiles (top to bottom; S/N(H$\alpha$) = 276, 104, 64, and 37).}\label{fig2} 
\end{figure*}

\section{Observations and data reduction}

NGC\,1365 was observed as part of the TYPHOON Program that uses the Progressive Integral Step Method (PrISM) also know as ``step-and-stare'' or ``stepped-slit'' technique to construct 3D data of nearby galaxies. Details about the TYPHOON Program will be presented in a forthcoming paper (Seibert et al. in preparation). Here, we provide a short summary, focusing on aspects immediately relevant to this study.  

NGC\,1365 was observed using the Wide Field reimaging CCD (WFCCD) imaging spectrograph on the 2.5-m du Pont telescope at the Las Campanas Observatory. The WFCCD has a field of view of $25\arcmin$ and we construct the 3D data cube using a custom long-slit ($18\arcmin\times1.65\arcsec$; 0.5 arcminute$^2$) which was placed along the north-south direction and progressively scanned across the galaxy through stepping and staring. Each pointing position was integrated for 600 seconds before the slit was moved by one slit width in the east-west direction for the next integration. This procedure was repeated until the optical disk of NGC\,1365 is covered. In total, 223 observations covering an area of approximately $6\arcmin\times18\arcmin$ were taken during 15 nights over four observing runs in November 2011, January 2016, February 2016 and August 2016. Observations were performed only when the seeing was less than the slit-width of $1.65\arcsec$. Spectrophotometric flux standards were observed each night.

The data are reduced using standard long-slit data reduction techniques. The wavelength calibration has a typical root-mean-square value 0.05\AA\  for the entire data set. Flux calibration is accurate to 2\% at the spaxel scale over the range of 4500 to 7500\AA. The reduced long-slit 2D spectra are tiled together to form the 3D data cube. The final reduced data cube covers a wavelength range of 3650 to 8150\AA, with spectral and spatial samplings of 1.5\AA\ and 1.65\arcsec, respectively. From fitting Gaussians to field stars in the reduced data cube, we estimate that the full-width half maximum of the point spread function is approximately 2\arcsec, which corresponds to approximately 175~pc at the assumed distance (Table~\ref{tbl1}). The instrumental dispersion is approximately 3.5\AA\ ($\sigma$; correspond to $\rm R\approx 850$ at 7000\AA).

The left panel of Figure~\ref{fig1} shows the BVR color-composite image extracted from the reduced 3D data cube. The middle and right panels show the H$\alpha$ flux and gas velocity maps derived from the data cube (see below for details).

\section{Data analysis}

To probe the oxygen abundance distribution in NGC\,1365, we first construct an H$\alpha$ emission line map from the reduced data cube. We then identify \HII\ regions in the H$\alpha$ map. The emission line fluxes of each \HII\ region are then measured and subsequently used to constrain the oxygen abundance and ionization parameter.

\subsection{Measure emission line fluxes}
We measure emission line fluxes using \lzifu\ \citep{Ho:2016rc,Ho:2016aa}. The emission line fitting tool \lzifu\ adopts \ppxf\ \citep{Cappellari:2004uq,Cappellari:2017aa} to model and subtract the stellar continuum, and the Levenberg-Marquardt least-squares method to fit emission lines as Gaussians \citep{Markwardt:2009lr}. We fit the stellar continuum using the {\scshape miuscat} simple stellar population models \citep{Vazdekis:2012yq} with 13 ages\footnote{Equally spaced on a logarithmic scale between 0.063 and 15.85Gyr.} and 3 metallicities\footnote{$\rm [M/H] =  -0.71,\ 0,\ +0.22$}. For the emission lines, we model 8 emission lines simultaneously as simple Gaussians (H$\beta$, [\ion{O}{3}]$\lambda\lambda4959,5007$,  [\ion{N}{2}]$\lambda\lambda6548,6583$, H$\alpha$, and  [\ion{S}{2}]$\lambda\lambda6716,6731$). We tie together the velocities and velocity dispersions of all the lines, and fix the flux ratios of [\ion{O}{3}]$\lambda\lambda4959/5007$ and [\ion{N}{2}]$\lambda\lambda6548/6583$ to those given by quantum mechanics. Hereafter, we will remove the wavelength subscription when appropriate, i.e. [\ion{O}{3}] $\equiv$ [\ion{O}{3}]$\lambda$5007,  [\ion{N}{2}] $\equiv$ [\ion{N}{2}]$\lambda$6583, and  [\ion{S}{2}] $\equiv$ [\ion{S}{2}]$\lambda$6716 +  [\ion{S}{2}]$\lambda$6731.

\begin{figure*}[!ht]
\centering
\includegraphics[width=0.75\textwidth]{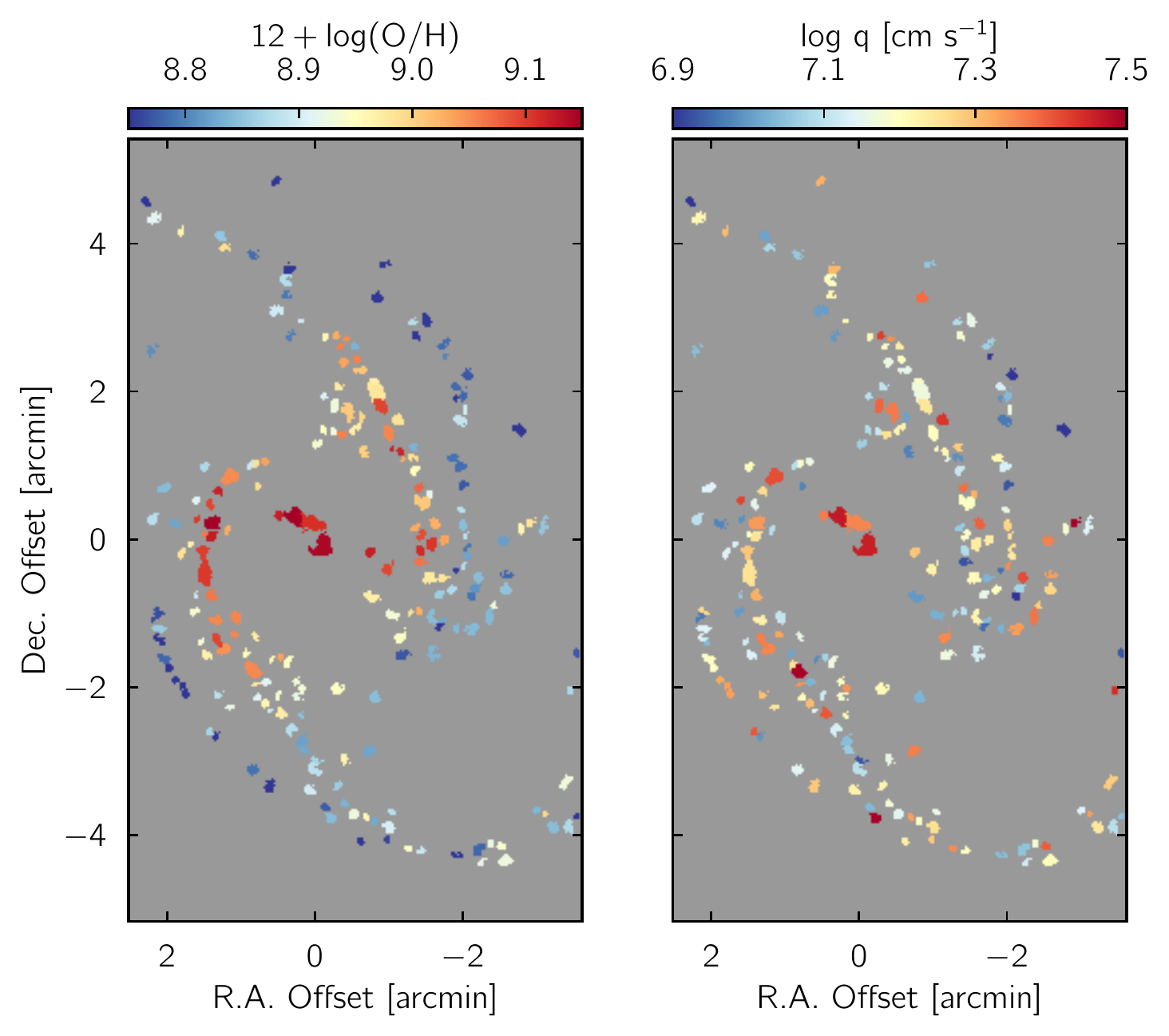}
\caption{Gas-phase oxygen abundance (left) and ionization parameter (right) maps. Each region on the maps correspond to one \HII\ region identified in the H$\alpha$ map in Figure~\ref{fig1}. We note that the abundance scale here is directly from the photoionization model of \citet{Dopita:2013qy}. The oxygen abundances measured from temperature sensitive auroral lines by \citet{Bresolin:2005uq} are approximately 0.7 solar. }\label{fig3}
\end{figure*}

The emission line fitting is first performed on every pixel to construct the H$\alpha$ map. From this H$\alpha$ map (Figure~\ref{fig1}), we identify \HII\ regions using an adapted version of the \idl\ routine \hiiphot\ developed by \citet{Thilker:2000rf}. \hiiphot\ identifies \HII\ regions on narrowband H$\alpha$ images using object recognition techniques that can accurately characterize the photometric properties of \HII\ regions, taking into account possible source irregularity. \hiiphot\ has been applied to narrow band images of nearby galaxies to characterize physical properties of \HII\ regions \citep[e.g. size distribution and luminosity function;][]{Thilker:2002uq,Helmboldt:2005fj,Helmboldt:2009kx}. The initial detection yields 308 \HII\ regions on our H$\alpha$ map, purely based on the H$\alpha$ surface brightness distribution in NGC\,1365.

To robustly measure the \HII\ region line fluxes from the data, we spatially bin our data cube based on the \HII\ region mask to produce synthetic  \HII\ region spectra. We then re-measure the emission line fluxes using the same procedure described above. We reject regions with signal-to-noise ratios (S/Ns) less than 3 in any of the diagnostic lines (H$\beta$, [\ion{O}{3}], [\ion{N}{2}], H$\alpha$, and [\ion{S}{2}]), and also those with line ratios inconsistent with photoionization based on their [\ion{N}{2}]/H$\alpha$ and [\ion{O}{3}]/H$\beta$ line ratios \citep[e.g.~excited by shocks or AGN;][]{Kewley:2001lr,Kauffmann:2003vn,Kewley:2006lr}. Based on these selection criteria, 212 \HII\ regions enter our analysis. Figure~\ref{fig2} demonstrates the quality of our \HII\ region spectra. The four example spectra are selected from the first to the last quartiles based on S/N(H$\alpha$).

We note that identifying \HII\ regions from seeing limited data is non-trivial. It is particularly difficult to separate \HII\ region complexes, especially toward the center of the galaxy due to source crowding. The major purpose of building an \HII\ region mask and synthesizing  \HII\ region spectra is to increase the S/N of the weak lines and avoid the line fluxes being contaminated by non-thermal emission (e.g.~diffuse ionized gas, shocked, AGN, etc.). It is possible that one ``\HII\ region'' defined in our analysis may contain multiple, poorly resolved \HII\ regions. In this case, the oxygen abundance and ionization parameter derived using the integrated line fluxes are approximately the luminosity-weighted mean oxygen abundance and ionization parameter.

\subsection{Oxygen abundance and ionization parameter}
We constrain the oxygen abundance and ionization parameter of each \HII\ region using a grid of \HII\ region photoionization models. We first extinction correct each \HII\ region based on its Balmer decrement (i.e. H$\alpha$/H$\beta$ = 2.86 for case-B recombination), assuming the extinction curve parameterized by \citet[][assuming $R_V=3.1$]{Cardelli:1989qy}. 

Our \HII\ region photoionization grid that forms the basis of our analysis are the latest \mappingsiv\ photoionization models from \citet[][D13]{Dopita:2013qy}. Specifically, these models use \textsc{starburst99} stellar atmospheres \citep{Leitherer:1999fk} and solar abundances of the ISM from \citet{Grevesse:2010vn}, and assume a pressure of $P/k=10^5$ ($n\sim 10~\rm cm^{-3}$) and that the electrons follow a Maxewell-Boltzmann distribution ($\kappa=\infty$). For full details see D13.

\begin{figure*}[!ht]
\centering
\includegraphics[width=\textwidth]{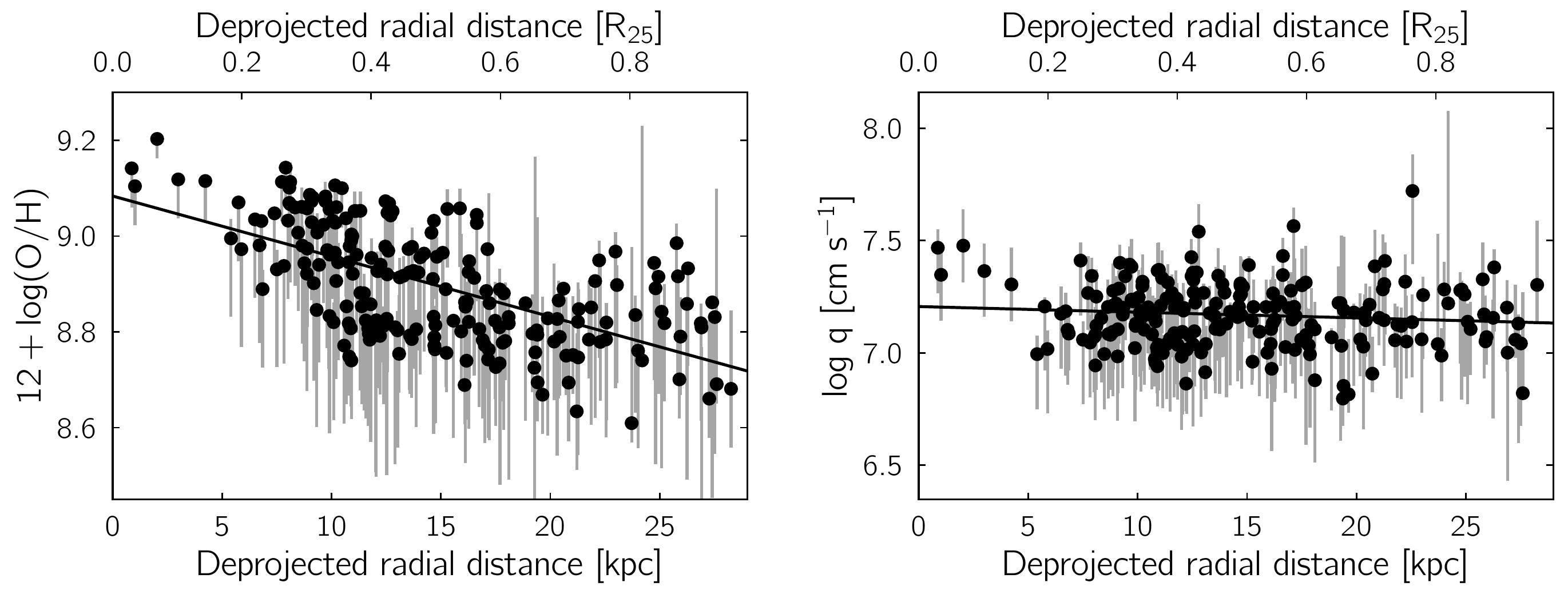}
\caption{Oxygen abundance (left panels) and ionization parameter (right panels) gradients in NGC\,1365. Each color point corresponds to one \HII\ region in Figure~\ref{fig3}. The black solid lines are the linear best fits. The negative abundance gradients are well defined, but the ionization parameter gradients are consistent with being flat. The best-fit lines are parametrized in Table~\ref{tbl2}.}\label{fig4}
\end{figure*}

To map the measured \HII\ region line emission onto the photoionization model grid and determine the physical parameters, we employ the Bayesian inference code \izi\ \citep{Blanc:2015qy}. \izi\ computes the posterior probability distributions given the grid and line fluxes. We report here the marginalized mean oxygen abundances and
ionization parameters calculated by \izi, assuming flat prior probability distributions for both parameters, bounded by the limits of the photoionization grid.

We note that the oxygen abundance can be derived from multiple diagnostics and calibrations in the literature. Different methods often yield discrepant oxygen abundances (see \citealt{Kewley:2008qy} and references therein). In general, the relative values of the oxygen abundance are robust on a statistical basis, but the absolute values can differ by over 0.3~dex. A common, practical approach is to convert the oxygen abundances derived from different calibrations to a common scale using empirical relationships, for example those determined by \citet{Kewley:2008qy}. This approach is not adopted in this work because we make no attempt to compare between results derived from multiple calibrations. In fact, the \HII\ region oxygen abundances derived from temperature sensitive auroral lines by \citet{Bresolin:2005uq} are approximately 0.7 solar, approximately 0.35~dex lower than those derived from the D13 grid. Although we only present the oxygen abundances derived from the D13 grid in the main body of the paper, we show in Appendix~A the oxygen abundances derived from other calibrations. We emphasize that the main results of this work are independent of the adopted diagnostics and calibrations.

\section{Results}

\subsection{Oxygen abundance and ionization parameter maps}
Figure~\ref{fig3} presents the oxygen abundance (in $\rm 12 + \log (O/H)$) and ionization parameter ($\log q$) maps of NGC\,1365. Overall, the oxygen abundance distribution in NGC\,1365 shows a clear radial gradient, with abundances that are on average higher at smaller radii than at larger radii. The ionization parameter map, however, shows no clear visible structure or radial gradient.

\subsection{Oxygen abundance and ionization parameter gradients}

We present radial profiles of the oxygen abundance and ionization parameter in Figure~\ref{fig4}. The geometric centers of the \HII\ regions are deprojected to the galactic disk frame using the parameters tabulated in Table~\ref{tbl1}, assuming a circular thin disk. To quantify the observed oxygen abundance and ionization parameter gradients, we perform unweighted least-squares linear fits. We fit the bootstrapped data 1,000 times and report the means and standard deviations of the best-fits. The best-fit gradients are shown in Figure~\ref{fig4} and the parameters summarized in Table~\ref{tbl2}. The negative oxygen abundance gradient in NGC\,1365 is consistent with those measured in other field galaxies in the local Universe \citep{Ho:2015hl}. The ionization parameters are consistent with presenting a flat gradient. 

We note that at small radii ($<5$kpc) there are indications that the oxygen abundances and ionization parameters are systematically higher than the linear fits. However, ``\HII\ regions'' defined here suffer from source crowding (see above). Their line ratios could also be contaminated by the central AGN and AGN-driven outflows \citep{Lena:2016ud}. It is possible this feature is an artifact.

\begin{figure}
\centering
\includegraphics[width=0.8\columnwidth]{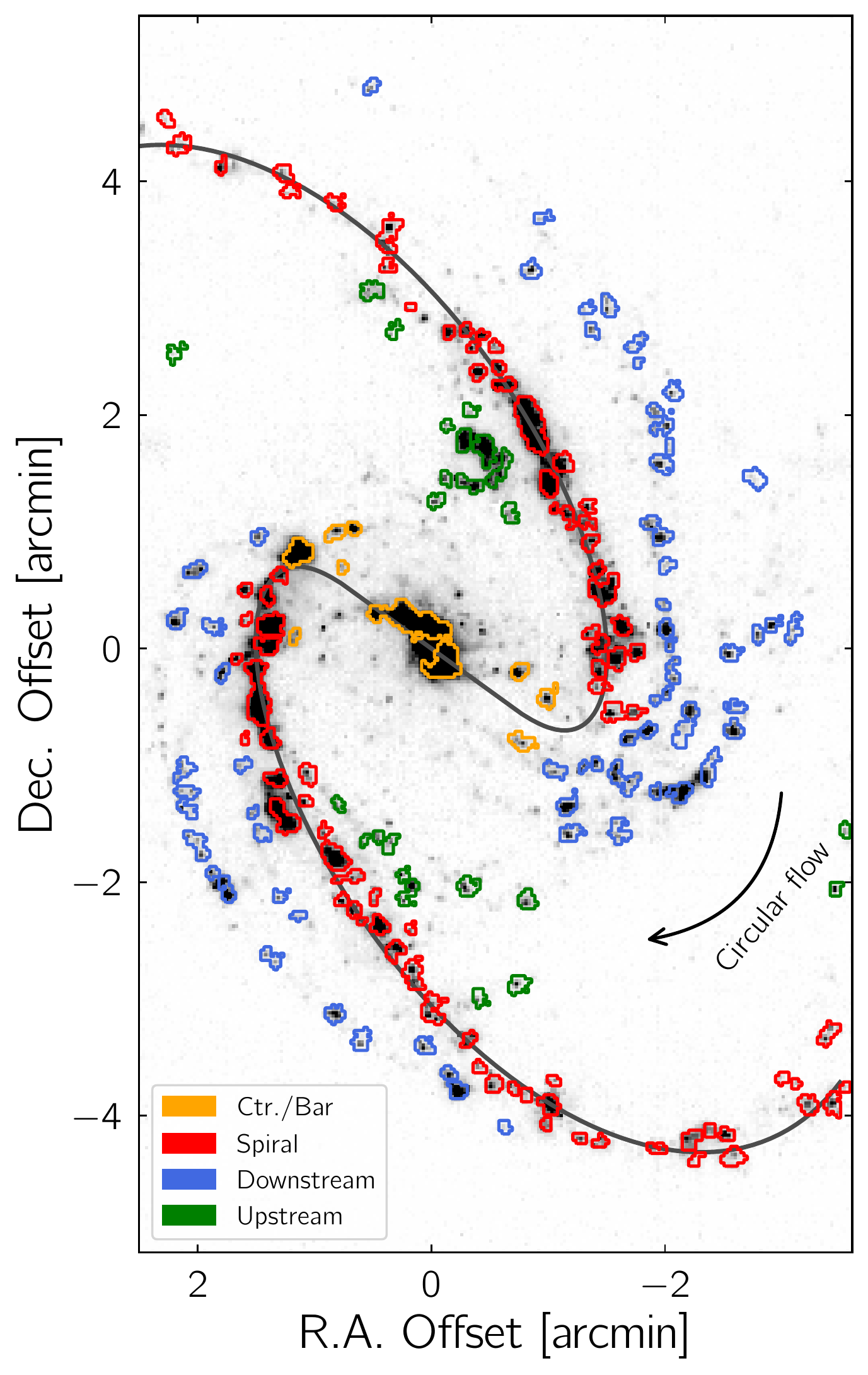}
\caption{Classification of \HII\ regions based on their positions relative to the spiral pattern. The background gray-scale shows the H$\alpha$ map presented Figure~\ref{fig1}. The solid curve indicates the spiral arms defined by fitting the analytic formula proposed by \citet{Ringermacher:2009oq} to bright \HII\ regions on the spiral arms. The colored contours outline the \HII\ regions identified by \hiiphot\ and presenting emission line ratios consistent with photoionization. The different colors correspond to the different groups of \HII\ regions described in Section~4.3.}\label{fig5}
\end{figure}

\begin{figure*}
\centering
\includegraphics[width=\textwidth]{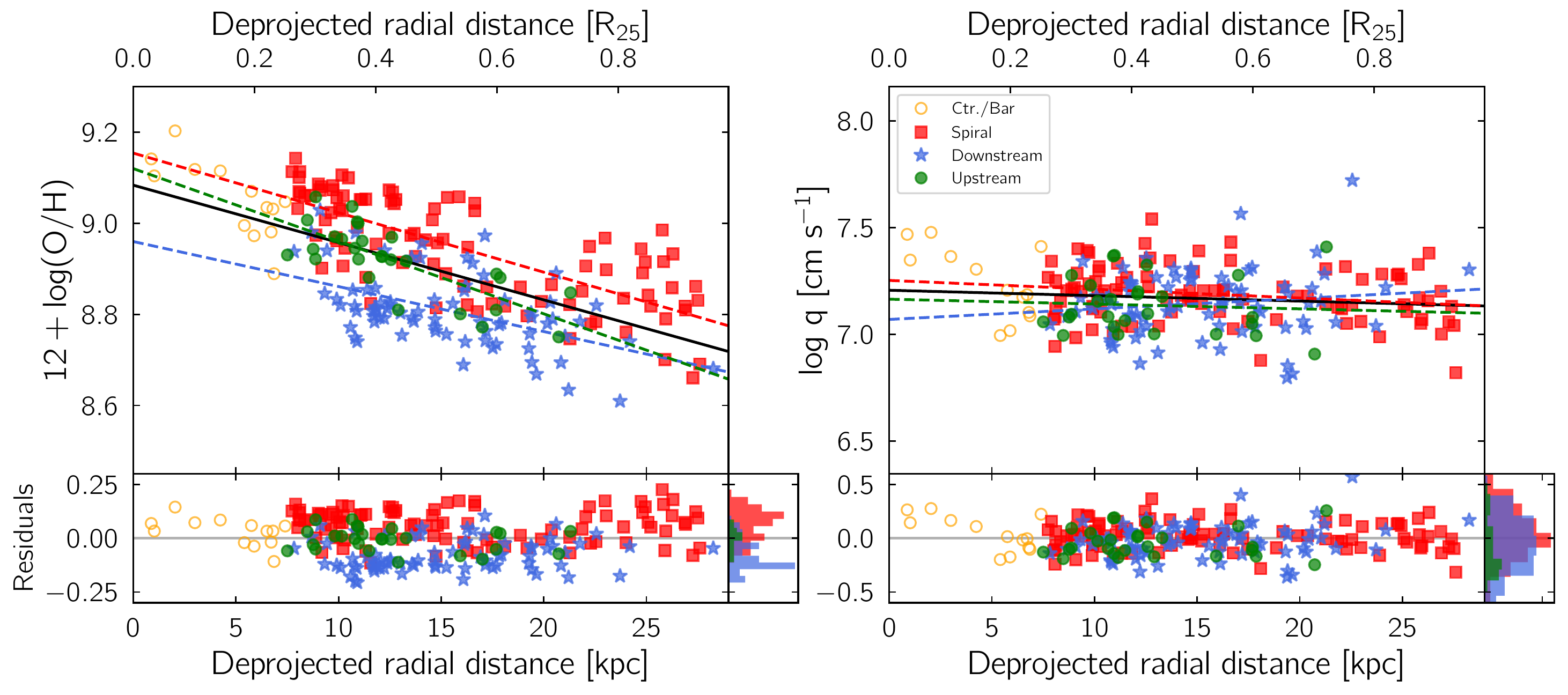}
\caption{Oxygen abundance (left panels) and ionization parameter (right panels) gradients in NGC\,1365. Each color point corresponds to one \HII\ region in Figure~\ref{fig3}. The different color symbols correspond to the different groups of \HII\ regions shown in Figure~\ref{fig5}. The black solid lines are the overall best fits. The red, blue and green dashed lines are the linear best fits to the ``Spiral'', ``Downstream'' and ``Upstream'' \HII\ regions, respectively. The bottom panels show the residuals, i.e.~after subtracting the black solid lines. The best-fit parameters are tabulated in Table~\ref{tbl2}.}\label{fig6}
\end{figure*}

\subsection{Azimuthal variation of oxygen abundance}

A more subtle, yet striking, feature in Figure~\ref{fig3} is the azimuthal variation of oxygen abundance. On the color composite image and H$\alpha$ map (Figure~\ref{fig1}) the two spiral arms in NGC\,1365 are well defined by the prominent young stars and bright \HII\ regions. A secondary set of \HII\ regions can be identified on the leading side of the main spiral arms. These ``downstream'' \HII\ regions are particularly obvious on the H$\alpha$ map as strings of dimmer yet continuously distributed \HII\ regions virtually parallel to the spiral arms. In Figure~\ref{fig3}, these downstream \HII\ regions appear to have, on average, lower oxygen abundances compared to regions on the spiral arms at similar radii. 

\begin{table}[!hb]
 \caption{Oxygen abundance and ionization parameter gradients}
 \label{tbl2}
\centering
 \begin{tabular}{cccc}
  \hline
  Selection & Slope & Slope  & Offset \\
		 & [$\rm dex~kpc^{-1}$]  & [$\rm dex~R_{25}^{-1}$] & [dex] \\
  \hline
   \multicolumn{4}{c}{$\rm 12 + \log(O/H)$} \\
  \hline
  All        & $(-1.25\pm0.12)\times10^{-2}$ & $-0.37\pm0.04$ & $9.083\pm0.018$ \\
  Spiral     & $(-1.32\pm0.12)\times10^{-2}$ & $-0.39\pm0.04$ & $9.155\pm0.019$ \\
  Downstream & $(-0.99\pm0.19)\times10^{-2}$ & $-0.29\pm0.06$ & $8.960\pm0.030$ \\
  Upstream   & $(-1.61\pm0.29)\times10^{-2}$ & $-0.48\pm0.09$ & $9.122\pm0.038$ \\
   \hline
   \multicolumn{4}{c}{$\log q\ [\rm cm~s^{-1}]$} \\
   \hline
  All        & $(-2.58\pm1.85)\times10^{-3}$ & $-0.08\pm0.05$ & $7.207\pm0.028$ \\
  Spiral     & $(-3.97\pm2.04)\times10^{-3}$ & $-0.12\pm0.06$ & $7.250\pm0.035$ \\
  Downstream & $(+5.18\pm4.64)\times10^{-3}$ & $ 0.15\pm0.14$ & $7.067\pm0.066$\\
  Upstream   & $(-1.87\pm9.32)\times10^{-3}$ & $-0.06\pm0.28$ & $7.161\pm0.112$\\
  \hline
 \end{tabular}
\end{table}

To quantify the azimuthal variation, we first define where the main spiral arms are. We fit the analytic formula proposed by \citet{Ringermacher:2009oq} to bright \HII\ regions on the main spiral arms which we identified by eye. The analytic formula is designed to quantitatively characterize spiral galaxies along the Hubble sequence, including grand design and barred spirals. There are six parameters in the formula, including three parameters ($N, B, A$) describing the shape of the spiral and bar patterns in the frame of the galaxy, and three Euler angles ($\alpha, \beta, \gamma$) describing the projection from the frame of the galaxy to the plane of the sky. We use the Levenberg-Marquardt algorithm to minimize the cost function that is defined to be the sum of squares of the distances from individual \HII\ regions to the spiral arms. We adopt the parameters in Figure~3 of \citet{Ringermacher:2009oq} as the initial guess and arrive at the best fit values $(N, B, A, \alpha, \beta, \gamma) = (6.66, 0.99, 3.67\arcmin, 48.4^\circ, 59.2^\circ, 29.2^\circ)$. 

The best-fit spiral pattern is shown in Figure~\ref{fig5}. We categorize each \HII\ region based on its position relative to the best-fit spiral pattern. Within 1.5~kpc ($=0.3\arcmin$) of the spiral pattern, the \HII\ regions belong to the main spiral arms (``spiral'' in Figure~\ref{fig5}). Those regions within the central 7~kpc in the frame of the galactic disk (approximately where the best-fit spiral arms begin) are labeled as ``central/bar''. The remaining regions are classified as ``downstream'' or ``upstream'' based on their positions relative to the spiral arms. We will focus on discussing the ``spiral'', ``upstream'' and ``downstream'' regions in the rest of the paper.

In Figure~\ref{fig6}, we present again the oxygen abundance and ionization parameter gradients and color-code the data points based on the different groups defined in Figure~\ref{fig5}. Figure~\ref{fig6} further demonstrates the azimuthal variation visible on the oxygen abundance map (Figure~\ref{fig3}). The \HII\ regions that belong to the spiral arms have systematically higher oxygen abundances than the other regions at the same galactocentric distances. The upstream regions have oxygen abundances {\it in between} the main spiral and downstream \HII\ regions. A total of 73 of the 96 (76\%) main spiral \HII\ regions are above the best-fit abundance gradient. In contrast, only 13/76 (17\%) of the downstream \HII\ regions are above the best-fit abundance gradient. The distributions of the residuals, after the best-fit radial gradient is subtracted, are statistically different (Figure~\ref{fig6} histogram). Two-sample Kolmogorov–Smirnov tests suggest  that the null hypotheses that the residuals of the spiral-downstream, spiral-upstream, and downstream-upstream regions are drawn from the same distributions can effectively be ruled out ($p-{\rm value}=1.8\times 10^{-16}$, $1.8\times 10^{-5}$, and $3.2\times 10^{-5}$, respectively)

We also perform the same least-squares linear fits to the spiral arm, downstream and upstream \HII\ regions. The best-fit oxygen abundance gradients shown in Figure~\ref{fig6} demonstrate the systematic differences between the different groups of \HII\ regions. The best-fit parameters are tabulated in Table~\ref{tbl2}. It is worth pointing out that the majority of the upstream \HII\ regions (21/27; 78\%) are {\it in between} the best-fit gradients defined by the spiral arm and downstream \HII\ regions.

\begin{figure}
\centering
\includegraphics[width=0.8\columnwidth]{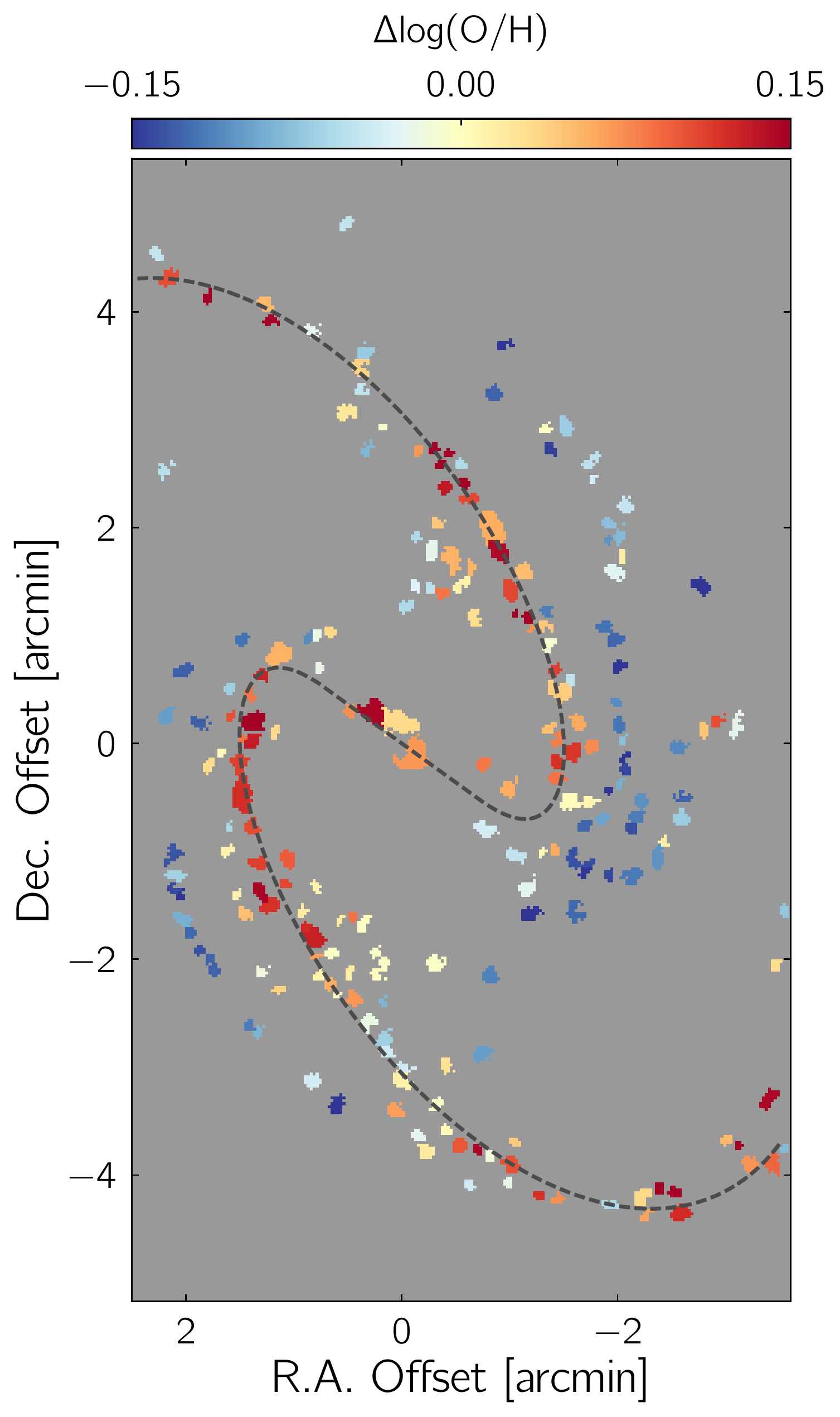}
\caption{Oxygen abundance map {\it after} subtracting the best-fit radial abundance gradient (i.e.~Figure~\ref{fig6}). The azimuthal variations of oxygen abundance are now clearly visible. }\label{fig7}
\end{figure}

In Figure~\ref{fig7}, we present the azimuthal variations more clearly by subtracting the best-fit radial gradient from the oxygen abundance map. If there are no azimuthal variations, the residual map should present only random fluctuations. However, Figure~\ref{fig7} shows that the residuals correlate strongly with the spiral pattern. 

To summarize, while overall there is a negative oxygen abundance gradient in NGC\,1365, we find pronounced azimuthal variations. The \HII\ regions on the spiral arms present, on average, about 0.2~dex higher oxygen abundance than the downstream \HII\ regions. The upstream \HII\ regions have abundances in between the main spiral arm and downstream \HII\ regions.

\section{Discussion}

\subsection{What causes the azimuthal variation of oxygen abundance?}
\begin{figure*}
\centering
\includegraphics[width=0.7\textwidth]{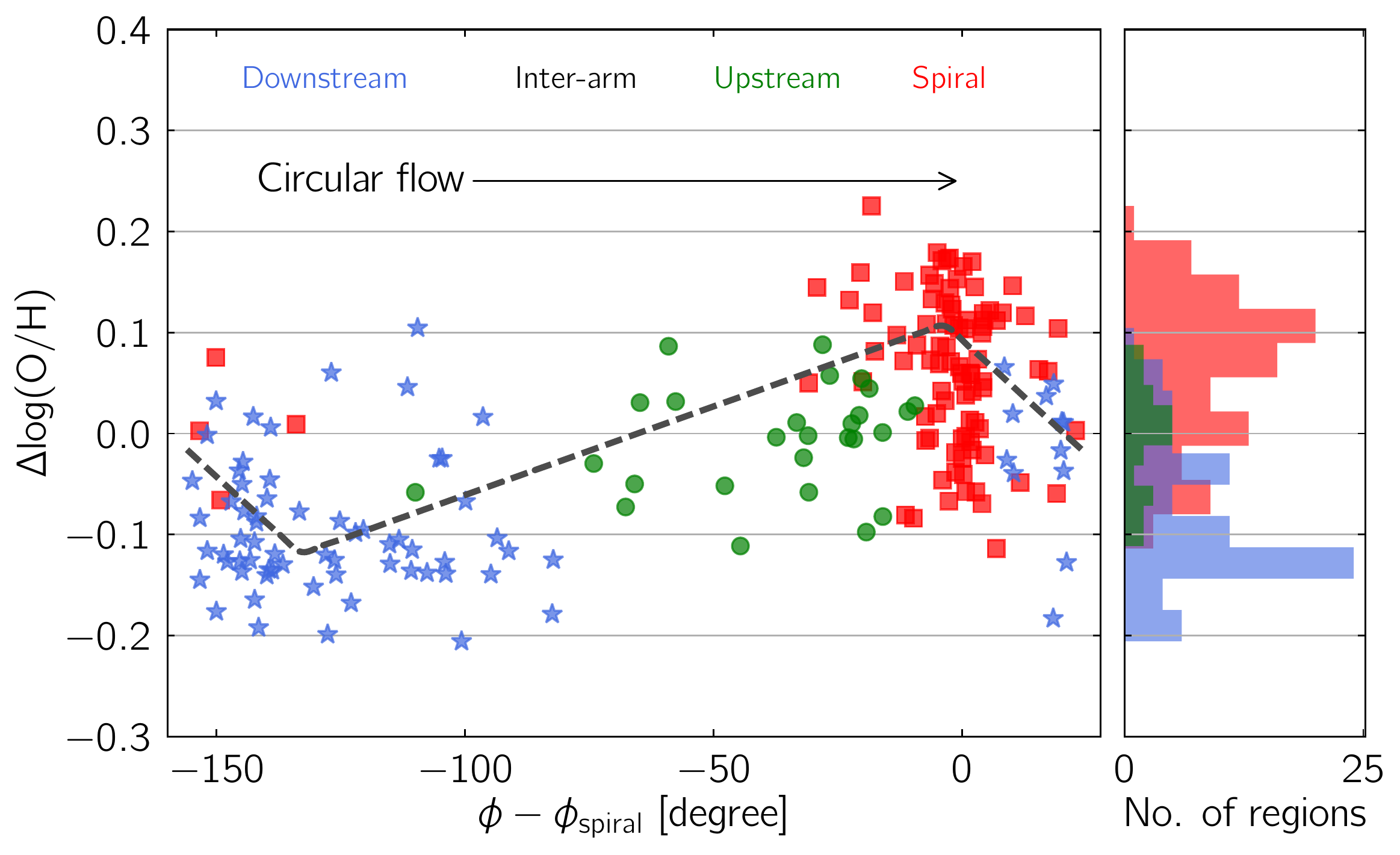}
\caption{{\it Left:} Oxygen abundance residuals versus phase angle relative to the spiral arms in the plane of the galaxy. The two spiral arms are matched in their phase angles and aligned to $\phi-\phi_{\rm spiral} = 0$. Increase in the phase angle corresponds to the clockwise direction in the 2D maps, the same direction as the circular gas flows. The dashed line is the best-fit asymmetric triangle wave function. {\it Right:} Histograms of the oxygen abundance residuals.}\label{fig8}
\end{figure*}

The azimuthal variations of oxygen abundance correlate strongly with the spiral pattern, with the positive abundance residuals peaking on the spiral arms. This implies that the physical processes driving the azimuthal variations are related to the dynamics of the gas and spiral pattern. 

In Figure~\ref{fig8}, we unwrap the 2D spatial abundance variations. We first calculate the azimuthal angle difference between each \HII\ region and the spiral arms (at the same radial distance), $\phi-\phi_{\rm spiral}$. We then match and align the two spiral arms such that the two arms are both at $\phi-\phi_{\rm spiral}=0$ and the phase angle spans 180 degrees. The gas orbital motion in NGC\,1365 flows in the clockwise direction, corresponding to increasing $\phi-\phi_{\rm spiral}$ in Figure~\ref{fig8} (see below).

If the pattern in Figure~\ref{fig8} persists over several orbital periods, then Figure~\ref{fig8} suggests that there are two different phases when a parcel of gas orbits around the galaxy. In the first phase, the gas parcel travels from downstream, through the inter-arm region, reaches the upstream and enters the spiral arm. During this period, the oxygen abundance increases over time. In the second phase, the gas parcel goes from the spiral arm to the downstream region, and experiences a sudden drop in the oxygen abundance. After the second phase, the gas parcel starts another rotation between phase 1 and phase 2.

In the first phase, the fact that the flow direction coincides with the direction of increasing oxygen abundance strongly suggests that self-enrichment is the main mechanism driving the azimuthal variations. The time available for self-enrichment  is approximately the dynamical time traveling through the inter-arm region, which depends on the rotation curve and the angular pattern speed of the spiral arms.

Although there is no direct estimate for the spiral pattern speed, there are several observational characteristics that suggest a likely range of values. The spiral, which appears to emanate from the ends of the bar, could be linked to the bar dynamics, for example. Different values of the bar co-rotation radius -- where the bar pattern speed is equal to the angular rotation rate of the disk material -- range from 100\arcsec\ to 150\arcsec\ (8.8 to 13.2~kpc), based on models of the \ion{H}{1} or ionized gas velocity fields \citep[][see table~5 in \citealt{Speights:2016uq} and references therein]{Teuben:1986fj,Lindblad:1996mz,Vera-Villamizar:2001nr,Zanmar-Sanchez:2008zr}.  The corresponding bar pattern speed ranges from 2.0 to 3.1 $\rm km~s^{-1}~arcsec^{-1}$ (22.8 to 35.3 $\rm~km~s^{-1}~kpc^{-1}$).  However, it seems unlikely that the spiral rotates with the same speed as the bar, as the majority of the observed  \HII\ regions at radii between 10 and 20~kpc lie on the convex side of the old stellar spiral arms, suggesting that they must lie inside the spiral co-rotation radius.  We thus favor scenarios in which the spiral rotates with its own, lower speed that is dynamically linked to that of the bar via mode-coupling \citep{Masset:1997rz}.  In either of the the two most commonly identified mode-coupling scenarios --  involving the overlap of the bar's co-rotation radius with either the inner Lindblad resonance of the spiral or its ultra-harmonic resonance -- the bar pattern speed noted above implies the spiral arm pattern speed is either $17\rm~km~s^{-1}~kpc^{-1}$ and $7\rm~km~s^{-1}~kpc^{-1}$, respectively \citep{Masset:1997rz}. This would place the spiral's co-rotation radius between 180\arcsec\ to 350\arcsec\ (16 to 31~kpc), and the majority of the \HII\ regions would lie inside this radius (60--100\%).  In this scenario, the region where we observe abundance variations sits between the co-rotation radii of the bar and spiral arms.  For the likely range of spiral pattern speeds, we estimate that the dynamical time to pass from one spiral arm to the next should be approximately 200 to 400~Myr.

This range of inter-arm dynamical times corresponds to approximately a few tens generations of massive stars. The oxygen produced by these stars, however, is not able to propagate far away from the production sites because the dynamical time is too short to effectively homogenize the chemical abundance via turbulent transport in a differentially rotating disk. \citet{Roy:1995ys} examine a variety of hydrodynamical processes and estimate that the mixing time-scale for 1 to 10~kpc scales to be around 1~Gyr. The mixing time for 0.1 to 1~kpc scales is much shorter ($\lesssim 10^8$~yr). \citet{de-Avillez:2002rw} adopt a three-dimensional, supernova-driven ISM model to estimate the time-scales for supernova explosions to erase different tracer fields in a $1\times1\times20$~kpc box. They found a mixing time of approximately 350~Myr for a Galactic supernova rate, with the mixing time decreasing with increasing supernova rate. Thus, we could expect chemical homogeneity only on sub-kpc scales. 

The chemical enrichment on sub-kpc scales could be enhanced by localized metal recycling, because succeeding star formation is likely to be triggered close to where the previous generations of massive stars have exploded. The virtually parallel spiral-like structures downstream from the spiral arms (particularly pronounced in the northern arm) are consistent with this hypothesis of triggered star formation. By recycling the enriched ISM, new generations of massive stars can reach progressively higher chemical abundances. The net effect is that the oxygen abundance in localized volumes contaminated by star formation increases over time \citep{Elmegreen:1998rf}. The rest of the ISM stays at roughly the same chemical abundance. Since the oxygen abundances we measured require the presence of \HII\ regions, we are only probing some fraction of the ISM that is being enriched continuously by star formation. This first, {\em self-enrichment phase} of rising oxygen abundance is likely driven by localized self-enrichment.

In the second phase, the oxygen abundance drops by about 0.2~dex from its peak on the spiral arms to the downstream regions. Dynamical effects by the density waves are likely to play a key role. When the gas enters the spiral density wave, molecular clouds close to the previous star formation sites (which are already quasi-unstable) can form stars more efficiently. Their higher metallicities facilitate the cooling and cloud collapse \citep{Kobayashi:2007fk}. The higher star formation rate surface density on the spiral arm increases the supernova rate, resulting in a shorter mixing time. \citet{de-Avillez:2002rw} show that the mixing time depends on the inverse square root of the supernova rate in their supernova-driven ISM simulations. In addition, frequent cloud-cloud collisions in the spiral arm also facilitate ISM mixing. The giant molecular cloud number density could increase by a factor of 20 or more in spiral arms, leading to a decrease in collision time-scale by the same factor \citep{Dobbs:2014ly}. Higher cloud-cloud collision speeds are also expected \citep{Fujimoto:2014gf}. We therefore expect a shorter mixing time-scale in the spiral arms than the inter-arm regions. Chemical homogeneity is possible on a spatial scale larger than that in the inter-arm region. 
This efficient mixing results in the dilution of more enriched gas by less enriched gas. This is because the oxygen produced by previous generations of stars and concentrated in and around active star formation sites can now be redistributed and mixes with the rest of the ISM. The measured \HII\ region oxygen abundances therefore drop in the second, {\em mixing-and-dilution phase}.

\subsection{A simple chemical evolution model}

To test if the picture described above can be quantitatively compared with the observations, we construct a simple, analytical chemical evolution model, based on \citet{Edmunds:1990fr}. 

First, we consider a spatial element with gas mass $m_g$, oxygen mass $m_o$, and oxygen abundance $Z$ ($\equiv m_o / m_g$). The spatial element starts from the downstream side and moves through the inter-arm region to the upstream and spiral arm regions. We denote the beginning and the end of this period as $t_0$ to $t_1$ with the subscripts 0 and 1, respectively. During this period, only a fraction of the gas in some small volume (with mass fraction $f$) is influenced by star formation (with the star formation rate $\psi$). Inside that small volume (with {\em mass fraction} $f$), the gas and oxygen are well-mixed in a short time-scale such that the instantaneous recycling assumption holds. The oxygen abundance inside that small volume increases over time (to $Z_{1,\psi}$), but the oxygen abundance of the rest of the gas remains the same ($Z_{1,\varnothing} = Z_{0}$). Since \HII\ regions form only in the small volume influenced by star formation, we only observe the abundance of the more enriched gas in the spatial element, but not the majority ($1-f$) of the gas that stays at the lower abundance, $Z_{1,\varnothing} = Z_{0}$.

After the spatial element reaches the spiral arm and experiences a burst of star formation, the mixing time-scale becomes short and thus the higher abundance gas enriched by star formation ($Z=Z_{1,\psi}$) mixes quickly with the gas uncontaminated by star formation ($Z=Z_{1,\varnothing} = Z_{0}$). This decreases the oxygen abundance and the overall oxygen abundance is now $Z=Z_{2}$. Here, we ignore the oxygen production from the burst of star formation and consider {\em only} gas mixing. This simplification is reasonable given that the time from $t_1$ to $t_2$ is short.

In this simple model, the period between $t_0$ and $t_1$ represents the self-enrichment phase described in the previous subsection, and $t_1$ to $t_2$ the mixing-and-dilution phase. Here, we assume that inflows and outflows play sub-dominant roles over the half rotation cycle considered, i.e. closed-box chemical evolution. 

If the gas mass and oxygen mass are $m_{g,0}$ and $m_{o,0}$ at $t=t_0$, then at $t=t_1$ the gas mass {\it inside} the volume influenced by star formation is
\begin{equation}
m_{g,1,\psi} = f m_{g,0} - \int_{t_0}^{t_1} (1-R) \psi(t) dt,
\end{equation}
and the oxygen mass is 
\begin{equation}
m_{o,1,\psi} = f m_{o,0} + \int_{t_0}^{t_1} y_o \psi(t) dt - \int_{t_0}^{t_1} (1-R) Z(t) \psi(t) dt.
\end{equation}
Here, $R$ is the returned mass fraction (i.e. fraction of the stellar mass which goes back to ISM through stellar winds) and $y_o$ is the oxygen yield. For the rest of the volume unaffected by star formation, the gas mass, oxygen mass, and oxygen abundance remain the same, i.e. 
\begin{equation}
m_{g,1,\varnothing}  =  (1-f) m_{g,0} ,
\end{equation}
\begin{equation}
m_{o,1,\varnothing}  =  (1-f) m_{o,0} ,
\end{equation}
and
\begin{equation}
Z_{o,1,\varnothing}  =  Z_{0} = m_{o,1,\varnothing} / m_{g,1,\varnothing} = m_{o,0} / m_{g,0}.
\end{equation}
At $t=t2$, the total gas mass inside the spatial element is then
\begin{equation}
m_{g,2} = (1-f) m_{g,0} + m_{g,1,\psi},
\end{equation}
and the total oxygen mass
\begin{equation}
m_{o,2} = (1-f) m_{o,0} + m_{o,1,\psi}.
\end{equation}
The difference in logarithmic abundance between $t_1$ and $t_0$ is
\begin{equation}\label{eq-deltaz10}
\log{Z_{1,\psi}} - \log{Z_0} = \log{ m_{o,1,\psi}/m_{g,1,\psi} \over m_{o,0}/m_{g,0} },
\end{equation}
between $t_1$ and $t_2$
\begin{equation}\label{eq-deltaz12}
\log{Z_{1,\psi}} - \log{Z_2} = \log{ m_{o,1,\psi}/m_{g,1,\psi} \over m_{o,2}/m_{g,2} },
\end{equation}
and between $t_0$ and $t_2$
\begin{equation}\label{eq-deltaz20}
\log{Z_{2}} - \log{Z_0} = \log{ m_{o,2}/m_{g,2} \over m_{o,0}/m_{g,0} }.
\end{equation}

To simplify the integrals in equations~1 and 2, we assume that $\psi(t)$ is constant, and the third right-hand-side term in equation~2, $Z(t)$, can is a constant over the period $t_0$ to $t_1$, i.e.~$Z(t)\approx{m_{o,0}/m_{g,0}}$. We discuss the second assumption in Appendix~B. Equations~\ref{eq-deltaz10}, \ref{eq-deltaz12} and \ref{eq-deltaz20} then become
\begin{equation}\label{eq-deltaz10-simple}
\log{Z_{1,\psi}} - \log{Z_0} = \log{\Big(1+ {{y_o \hat{\psi} \Delta t } \over 
f\hat{m}_{o,0} - (1-R) \hat{m}_{o,0} \hat{\psi} \Delta t}}\Big),
\end{equation} 
\begin{equation}\label{eq-deltaz12-simple}
\log{Z_{1,\psi}} - \log{Z_2}  = \log{\Bigg( { \hat{m_{o,0}} + {y_o \hat{\psi} \Delta t \over f - (1-R) \hat{\psi} \Delta t} \over
\hat{m_{o,0}} + {y_o \hat{\psi} \Delta t \over 1 - (1-R) \hat{\psi} \Delta t}} \Bigg)}, 
\end{equation}
and 
\begin{equation}\label{eq-deltaz20-simple}
\log{Z_2} - \log{Z_0}  = \log{\Big( 1 + { y_o \hat{\psi} \Delta t \over 
\hat{m}_{o,0} - (1-R) \hat{m}_{o,0} \hat{\psi} \Delta t  }  \Big) }.
\end{equation}
where $\hat{m}_{o,0} = m_{o,0}/m_{g,0} = Z_0$, $\hat{\psi} = \psi/m_{g,0}$ and $\Delta t = t_1 - t_0$. 

To compare equations \ref{eq-deltaz10-simple} to \ref{eq-deltaz20-simple} to our observation, we need to know $m_o$, $m_g$, and $\psi$. Unfortunately, large area CO mapping for the molecular gas disk has not been conducted yet. \citet{Tabatabaei:2013rf} used the 870$\mu m$ dust continuum emission to estimate the molecular gas mass of the disk. They estimated a total gas mass $M_g$ of $15.1\times10^9~M_\odot$ (\ion{H}{1} + $\rm H_{2}$), and a total SFR of the disk of 7.0 $M_\odot\rm~yr^{-1}$ (from FUV) and 16.7 $M_\odot\rm~yr^{-1}$ (from total infra-red luminosity). 

We assume that the spatial element at $t=t_0$ has $\psi/m_g$ and $m_o/m_g$ similar to the global average of the disk, i.e. 
\begin{equation}
\psi/m_{g,0} = \hat{\psi} \approx {12 M_\odot\rm~yr^{-1} \over 15.1\times10^9~M_\odot }
\end{equation}
and 
\begin{equation}
m_{o,0}/m_{g,0} = Z_0 = \hat{m}_{o,0} \approx 0.004.
\end{equation}
The former corresponds to a depletion time of $t_{\rm dep} = m_{g,0}/\psi = 1.26\rm~Gyr$, comparable to those measured in other nearby spiral galaxies \citep{Bigiel:2008yq,Leroy:2013rf}. The latter corresponds to the oxygen abundance of approximately 0.7 solar, as measured by \citet{Bresolin:2005uq} from \HII\ regions in NGC\,1365 using the direct temperature method. 

We adopt an oxygen yield $y_o$ of 0.00313 and a return mass fraction $R$ of 0.4. The oxygen yield is constrained empirically by \citet{Kudritzki:2015fp} to reproduce the Cepheids metallicity gradient in the Milky Way. For $\Delta t$, we use the dynamical time traveling between spiral arms estimated in the previous subsection, i.e. $\Delta t$ of 300~Myr. Assuming the mass fraction $f$ of 0.4, we reach 
\begin{equation}
\log{Z_{1,\psi}} - \log{Z_0}  = 0.23,
\end{equation}
\begin{equation}
\log{Z_{1,\psi}} - \log{Z_2}  = 0.15,
\end{equation}
and
\begin{equation}
\log{Z_2} - \log{Z_0}  = 0.08.
\end{equation}
These results are in agreement with our findings that the downstream \HII\ regions on average have approximately 0.2~dex lower oxygen abundance than the spiral arm \HII\ regions (e.g. Figures~\ref{fig7} and \ref{fig8}). 

We note that the assumed $f$ can be compared with the fact that approximately 10 to 30 percent of the disk area in typical star-forming galaxies is occupied by \HII\ regions at any given time, while the rest by the diffuse ionized gas \citep{Roy:1995ys,Zurita:2000sh}. The calculation above only considers a set of fiducial numbers for $y_o$, $R$, $\Delta t$ and $f$, each of which may carry some systematic uncertainties. In Appendix~B, we investigate further the model behavior when these quantities deviate from the fiducial values.

Our toy model is certainly too simple to fully describe reality. However,  by adopting reasonable guesses for the model parameters and mean physical properties (i.e.~star formation rate, gas content, and oxygen abundance) we can already reproduce the observed abundance variations to within 0.1~dex.  This suggests that the key underlying physical processes are already captured by the two-phase toy model.

\subsection{The chemical evolution carousel of spiral galaxies}

\begin{figure}
\centering
\includegraphics[width=0.9\columnwidth]{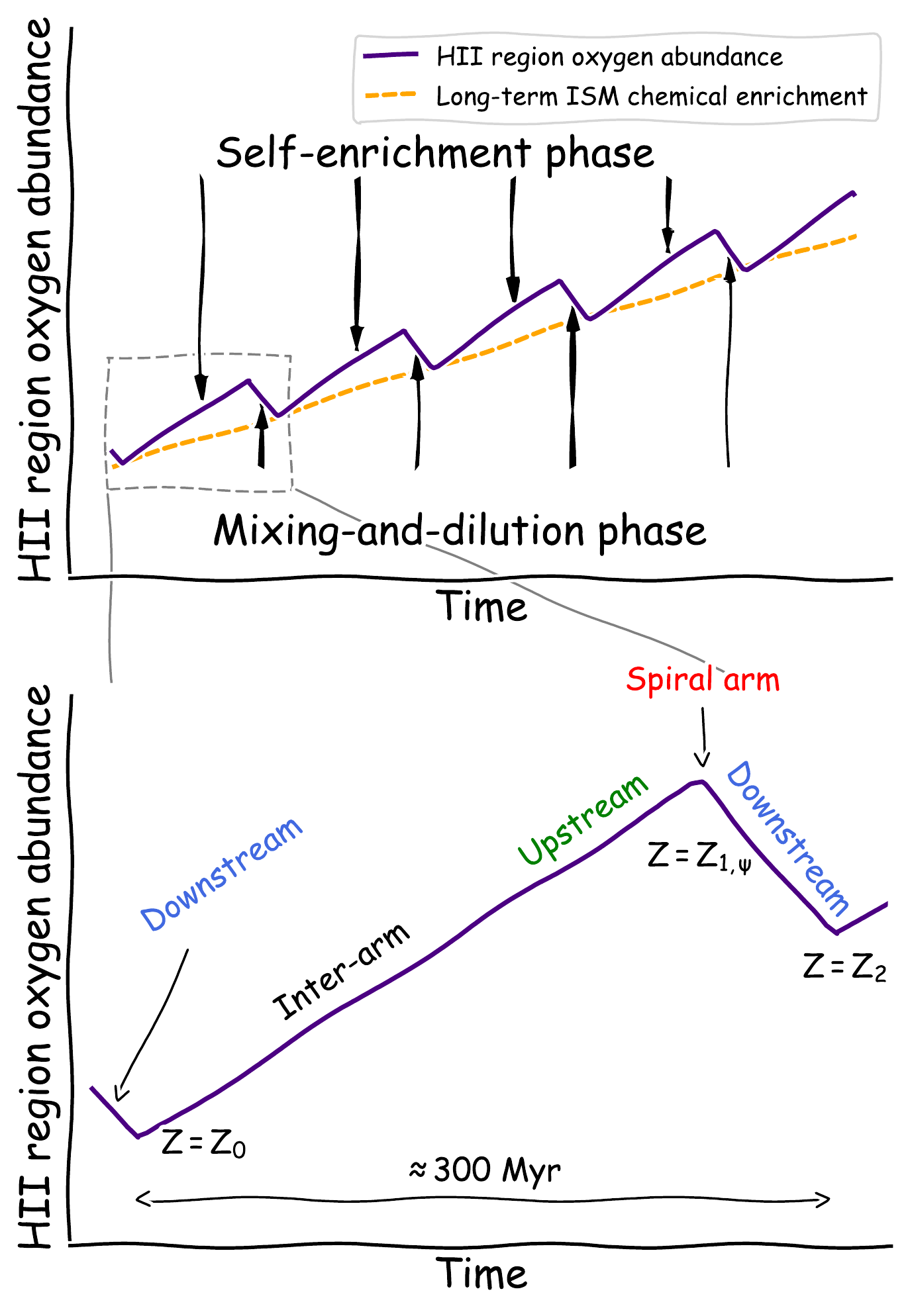}
\caption{Schematics illustrating the two different phases gas experiences when orbiting circularly in a spiral galaxy. The figure is not to scale. }\label{fig9}
\end{figure}

We summarize the picture described above in Figure~\ref{fig9}. When a spatial element orbits around a galaxy, the long-term evolution of ISM oxygen abundance over multiple orbital periods is determined by star formation that drives the oxygen production. In the absence of inflows and outflows (i.e.~virtually closed-box evolution), the oxygen abundance increases monotonically with time. However, the short-term, half orbital cycle evolution is also controlled by how quickly gas can mix. When gas travels through the inter-arm region, inefficient mixing results in localized metal recycling, leading to the rapid increase of oxygen abundance in the vicinities of \HII\ regions. Efficient mixing due to passing of a spiral density wave dilutes and homogenizes the localized over-abundance, and thus the measured \HII\ region oxygen abundance decreases. The rotation of the self-enrichment and mixing-and-dilution phases leaves an imprint on the long-term evolution of ISM oxygen abundance in the form of the observed azimuthal variations. The zigzag pattern of oxygen abundance in the time domain is a natural consequence of the two-phase cycling, which can be qualitatively reproduced with more realistic chemical evolution models \citep[e.g.][]{Kobayashi:2011aa}.

In this picture, localized metal recycling in the inter-arm regions is one of the keys to producing large azimuthal variations that can be observed more easily. For the same star formation rate and gas content, galaxies with longer inter-arm dynamical times are expected to experience more generations of star formation and thus build up larger azimuthal variations of oxygen abundance. This prediction can be tested by observing galaxies with different spiral arm pitch angles, pattern speeds and numbers of spiral arms.

\subsection{The role of radial gas flows}

In the picture described above, radial gas flows are assumed to play a sub-dominant role. However, large-scale streaming motions are known to affect the spatial distribution of abundances . \citet{Grand:2016hb} use a high-resolution cosmological zoom simulation of a Milky Way-sized halo to demonstrate that radial flows associated with the spiral arm can produce an overdensity of metal-rich stars on the trailing side (upstream) of the spiral arm and an overdensity of metal-poor star on the leading side (downstream). This is a result of the flows offsetting the existing negative metallicity gradient in the radial direction. The gas flows radially outward at the trailing side (upstream) and inward at the leading side (downstream) of the spiral arm. This systematic motion yields, at a fixed radius, a metal-rich trailing edge (downstream), a metal-poor leading edge (upstream), and an intermediate metallicity on the spiral arm. Such a signature has been observed by \citet{Sanchez-Menguiano:2016zr} in NGC\,6754 using MUSE. \citeauthor{Sanchez-Menguiano:2016zr} find that the trailing side has approximately 0.1~dex (peak-to-peak) higher \HII\ region oxygen abundance than the leading side, consistent with the amplitude seen in the simulation by \citet{Grand:2015la}.

The signature reported by \citeauthor{Sanchez-Menguiano:2016zr} is not observed in NGC\,1365. We find that the oxygen abundances on the spiral arms are higher than {\em both} the leading (downstream) and trailing (upstream) sides of the spiral arms. In addition, the oxygen variations we observe are larger (0.4~dex; peak-to-peak) than theirs. If we assume that a radially inward flow persists over 150~Myr (half of the inter-arm traveling time) with a speed of 20 $\rm km~s^{-1}$ \citep{Schmidt:2016fj}, the flow can travel approximately 3.1~kpc and decrease the oxygen abundance by 0.04~dex (based on the oxygen abundance gradients in Table~\ref{tbl2}). It is unclear whether the flow rates and durations of the radial flows could be affected by different accretion histories that govern gas infalling onto the disk. Under the simplest assumptions,  the 0.04~dex variation is not enough to explain our observation but is in a better agreement with \citet{Sanchez-Menguiano:2016zr}. The mechanisms shaping the abundance variations in NGC\,1365 and NGC\,6754 are thus likely to be different.

\section{Summary}
We have investigated the oxygen abundance distribution of the ISM in the spiral galaxy NGC\,1365 using a new 3D data cube constructed in the TYPHOON Program. 
About 200 \HII\ regions were identified from the H$\alpha$ map extracted from the  data cube. We derived their oxygen abundances and ionization parameters using flux ratios of strong emission lines. We found the following.

\begin{itemize}

\item There is a well-defined negative oxygen abundance gradient in NGC\,1365, with a slope consistent with other local star-forming galaxies. The ionization parameters are consistent with a flat gradient. 

\item In addition to the radial dependency of oxygen abundance, NGC\,1365 presents pronounced, 0.2~dex azimuthal variations of oxygen abundance. The variations correlate systematically with the $m=2$ spiral arms in NGC\,1365. At a fixed radius, the oxygen abundances are highest on the spiral arms, lowest in the downstream side of the spiral arms, and intermediate in the upstream side. The coherent variations are seen in multiple abundance diagnostics and calibrations over the radial range of 8 to 20 kpc (0.3 to 0.7 $R_{25}$). 

\end{itemize}

We argue that the systematic azimuthal variations of \HII\ region oxygen abundance are caused by the following processes. 

When a parcel of gas travels from the downstream side of a spiral arm, through the inter-arm region, and reaches the upstream side of the next spiral arm, the oxygen produced by massive stars can be mixed only with the nearby gas. This is because the mixing time-scales on sub-kpc scale are shorter than the orbital time, but the mixing time-scales on kpc scale are longer. In the meantime, new generations of stars are also formed near where the previous generations of massive stars explode, i.e.~induced star formation. This results in localized metal recycling which builds up higher ISM oxygen abundances in small volumes within which \HII\ region oxygen abundances are measured. During this period, some of the ISM remains at their original abundances because they are not contaminated by star formation and mixing.

When the gas parcel enters the spiral arm, the spiral density wave triggers a burst of star formation in gas that is already quasi-unstable, i.e.~those gas that has high metallicities and is close to the previous star formation sites. The increase in star formation rate surface density and cloud-cloud collision frequency shortens the mixing time. Thus, chemical homogeneity can be established in volumes larger than those in the inter-arm region. Through efficient mixing, the previously more contaminated gas is diluted by less contaminated gas, resulting in the lower \HII\ region oxygen abundances measured in the downstream side of the spiral arm.

We have calculated a simple chemical evolution model to test if our picture could be quantitatively compared with the observation. We found that our toy model can reproduce the 0.2~dex azimuthal variations assuming reasonable model parameters. This suggests that the simple picture described above already captures the key physical processes driving the azimuthal variation of oxygen abundance in NGC\,1365.

\acknowledgments
This paper includes data obtained with the du Pont Telescope at the Las Campanas Observatory, Chile as part of the TYPHOON Program, which has been obtaining optical data cubes for the largest angular-sized galaxies in the southern hemisphere. We thank past and present Directors of The Observatories and the numerous time assignment committees for their generous and unfailing support of this long-term program.

We thank the anonymous referee for reviewing this paper and providing positive and constructive comments. ITH is grateful to the support of an MPIA fellowship. ITH and RPK acknowledge the support by the Munich Institute for Astro- and Particle Physics (MIAPP). BG gratefully acknowledges the support of the Australian Research Council as the recipient of a Future Fellowship (FT140101202). ES acknowledges funding from the European Research Council (ERC) under the European Union’s Horizon 2020 research and innovation programme (grant agreement No. 694343).

\facility{du Pont (WFCCD)}
\software{IDL, HIIphot, PPXF, MPFIT, LZIFU, IZI, Python, numpy, scipy, astropy, matplotlib}

\bibliography{references}

\appendix
\section{Oxygen abundance derived from other calibrations}

In the main text of the paper, we present the oxygen abundance derived by \izi\ using the D13 model grids from \mappingsiv. Here, we derive the oxygen abundance using three different strong-line calibrations. We derive the oxygen abundance using the O3N2 index first introduced by \citet{Alloin:1979uq}, 
\begin{equation}
{\rm O3N2} \equiv \log \Big({[\textrm{O}\textsc{iii}]\lambda 5007 / {\rm H\beta} \over [\textrm{N}\textsc{ii}]\lambda 6583/ {\rm H\alpha}}\Big). 
\end{equation}
Due to the close proximity in wavelength of the two line pairs, this index is insensitive to dust extinction. \citet[][P04]{Pettini:2004lr} and \citet[][M13]{Marino:2013vn} have calibrated the index to $T_e$ metallicity using high quality \HII\ region spectra. We adopt both calibrations to derive oxygen abundances. 

\begin{figure*}
\centering
\includegraphics[width=0.7\textwidth]{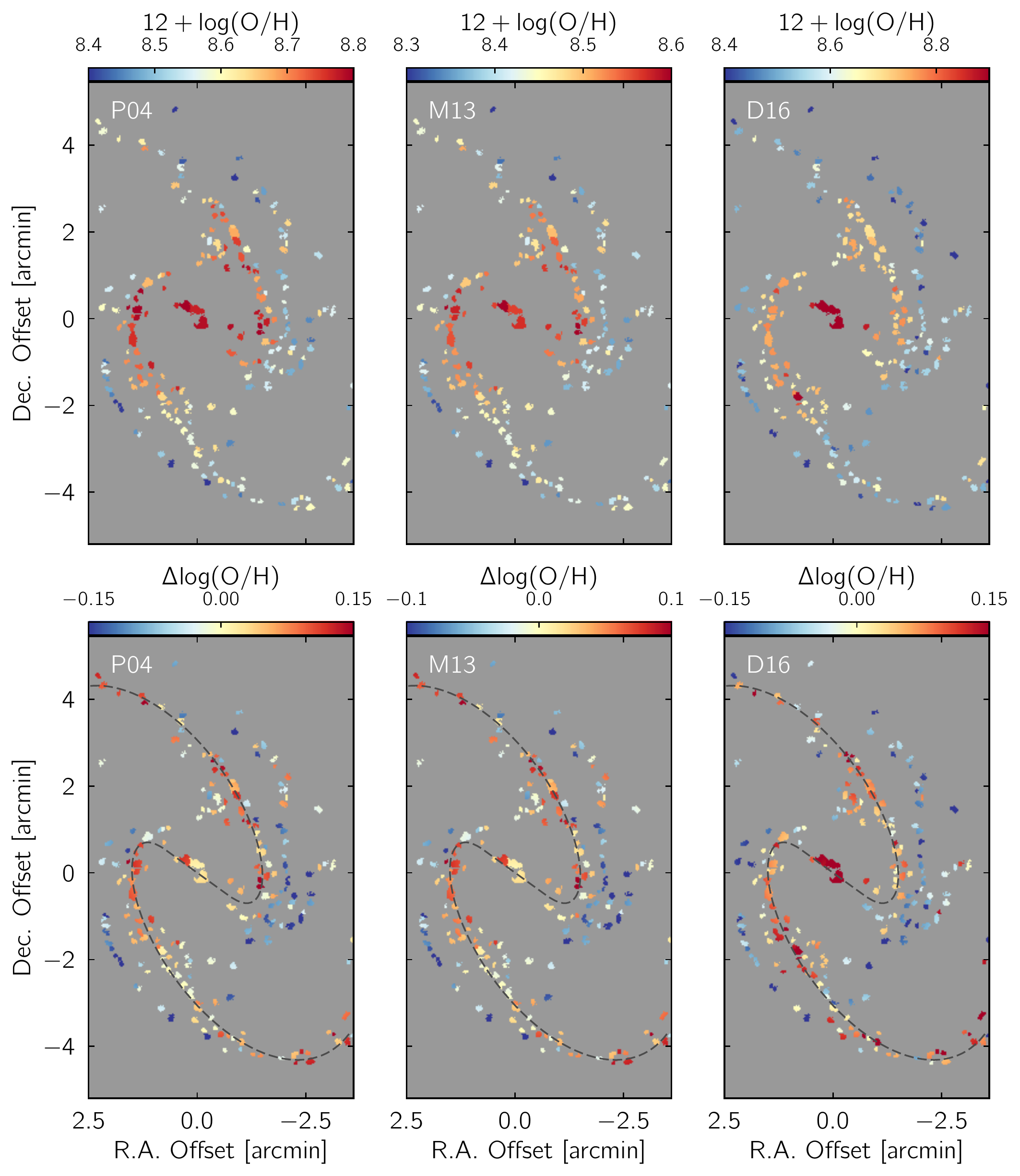}
\caption{Top row: oxygen abundances derived from different diagnostics and calibrations. From left to right: from the O3N2 diagnostic using the calibrations by \citet[][]{Pettini:2004lr} and \citet[][]{Marino:2013vn}, and from the N2S2 calibration by \citet[][]{Dopita:2016fk}. Bottom row: corresponding oxygen abundance residual maps after subtracting the best-fit abundance gradients (see Figure~\ref{fig7}). The dashed curves indicate the spiral pattern (see Figure~\ref{fig5}). The azimuthal variations of oxygen abundance are pronounced on all the maps. }\label{figa1}
\end{figure*}

\begin{figure*}
\centering
\includegraphics[width=\textwidth]{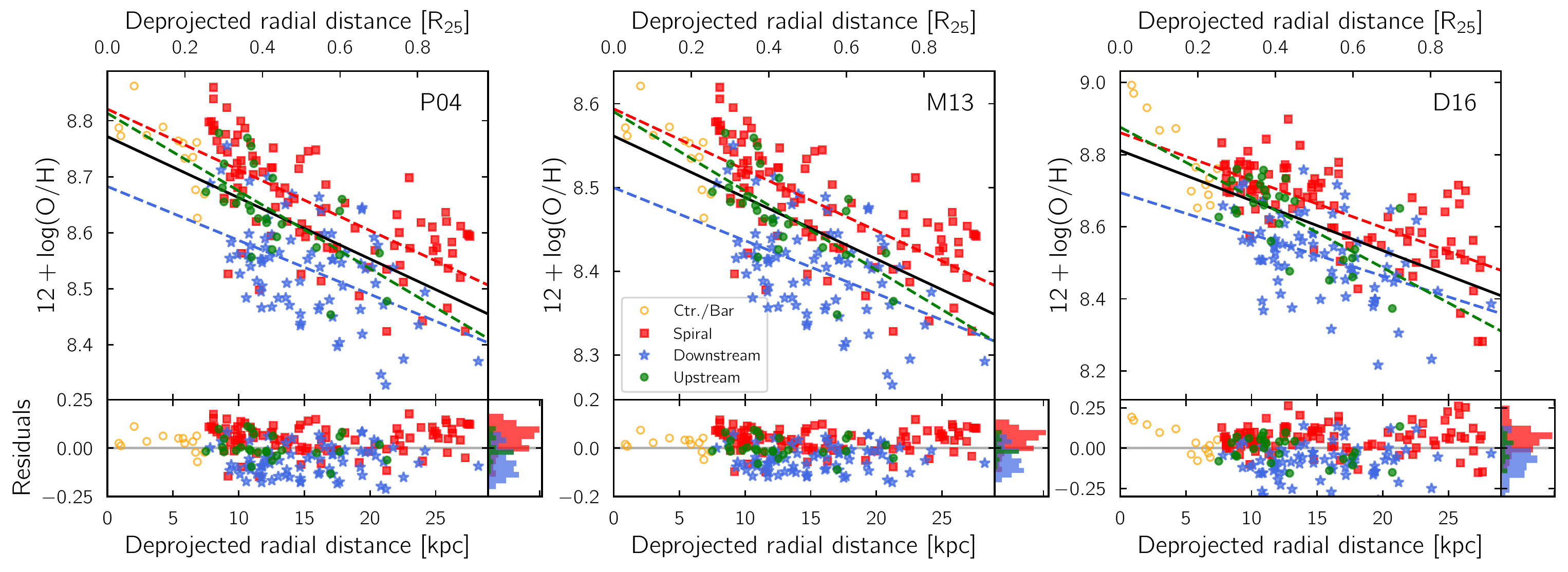}
\caption{Oxygen abundance gradients derived from different diagnostics and calibrations. The different color symbols correspond to the different groups of \HII\ regions shown in Figure~\ref{fig5}. The black solid lines are the linear best fits of all the \HII\ regions. The red, blue and green dashed lines are the best fits to the corresponding groups of \HII\ regions. The bottom panels show the residuals, i.e.~after subtracting the black solid lines ($\rm\Delta \log{(O/H)}$). The best-fit linear gradients are tabulated in Table~\ref{tbla1}.}\label{figa2}
\end{figure*}

\begin{table}
 \caption{Oxygen abundance gradient}
 \label{tbla1}
\centering
 \begin{tabular}{cccc}
  \hline
  Selection & Slope & Slope  & Offset \\
		 & [$\rm dex~kpc^{-1}$]  & [$\rm dex~R_{25}^{-1}$] & [dex] \\
  \hline
   \multicolumn{4}{c}{P04 $\rm 12 + \log(O/H)$} \\
  \hline
  All        & $(-1.09\pm0.10)\times10^{-2}$ & $-0.32\pm0.03$ & $8.771\pm0.015$ \\
  Spiral     & $(-1.08\pm0.12)\times10^{-2}$ & $-0.32\pm0.03$ & $8.820\pm0.020$ \\
  Downstream & $(-0.95\pm0.21)\times10^{-2}$ & $-0.28\pm0.06$ & $8.681\pm0.033$ \\
  Upstream   & $(-1.40\pm0.33)\times10^{-2}$ & $-0.41\pm0.10$ & $8.815\pm0.041$ \\
   \hline
   \multicolumn{4}{c}{M13 $\rm 12 + \log(O/H)$} \\
   \hline
  All        & $(-0.73\pm0.07)\times10^{-2}$ & $-0.22\pm0.02$ & $8.561\pm0.010$ \\
  Spiral     & $(-0.73\pm0.07)\times10^{-2}$ & $-0.21\pm0.02$ & $8.594\pm0.013$ \\
  Downstream & $(-0.64\pm0.14)\times10^{-2}$ & $-0.19\pm0.04$ & $8.501\pm0.021$ \\
  Upstream   & $(-0.96\pm0.23)\times10^{-2}$ & $-0.28\pm0.07$ & $8.592\pm0.028$ \\
  \hline
   \multicolumn{4}{c}{D16 $\rm 12 + \log(O/H)$} \\
   \hline
  All        & $(-1.39\pm0.15)\times10^{-2}$ & $-0.41\pm0.04$ & $8.812\pm0.021$ \\
  Spiral     & $(-1.32\pm0.15)\times10^{-2}$ & $-0.39\pm0.05$ & $8.861\pm0.022$ \\
  Downstream & $(-1.15\pm0.26)\times10^{-2}$ & $-0.34\pm0.08$ & $8.693\pm0.040$ \\
  Upstream   & $(-1.97\pm0.55)\times10^{-2}$ & $-0.58\pm0.16$ & $8.880\pm0.066$ \\
  \hline
\end{tabular}
\end{table}

We also calculate the oxygen abundance using the N2S2 diagnostic proposed by \citet[][D16]{Dopita:2016fk}. The diagnostic involves the [\ion{N}{2}]$\lambda$6583/[\ion{S}{2}]$\lambda\lambda$ 6716,6731 and [\ion{N}{2}]$\lambda$6583/H$\alpha$ line ratios that are also insensitive to dust extinction. D16 calibrate this diagnostic using the \mappingsv\ code and show that the derived oxygen abundance is insensitive to the change of ISM pressure.

The top row in Figure~\ref{figa1} presents the oxygen abundance maps derived from the three calibrations and demonstrates that the oxygen abundance gradients are pronounced in all the maps. The radial gradients are also shown in Figure~\ref{figa2}. In the bottom row of Figure~\ref{figa1} and also Figure~\ref{figa2}, we repeat the same analysis in Section~4.3, i.e.~subtracting the best-fit radial gradients. The residual maps and residuals all show clear azimuthal variations similar to those seen in Figures~\ref{fig6} and \ref{fig7}. We also tabulate the best-fit radial gradients in Table~\ref{tbla1}

\section{Chemical evolution model: deviation from the fiducial values}
\begin{figure*}
\centering
\includegraphics[width=0.7\textwidth]{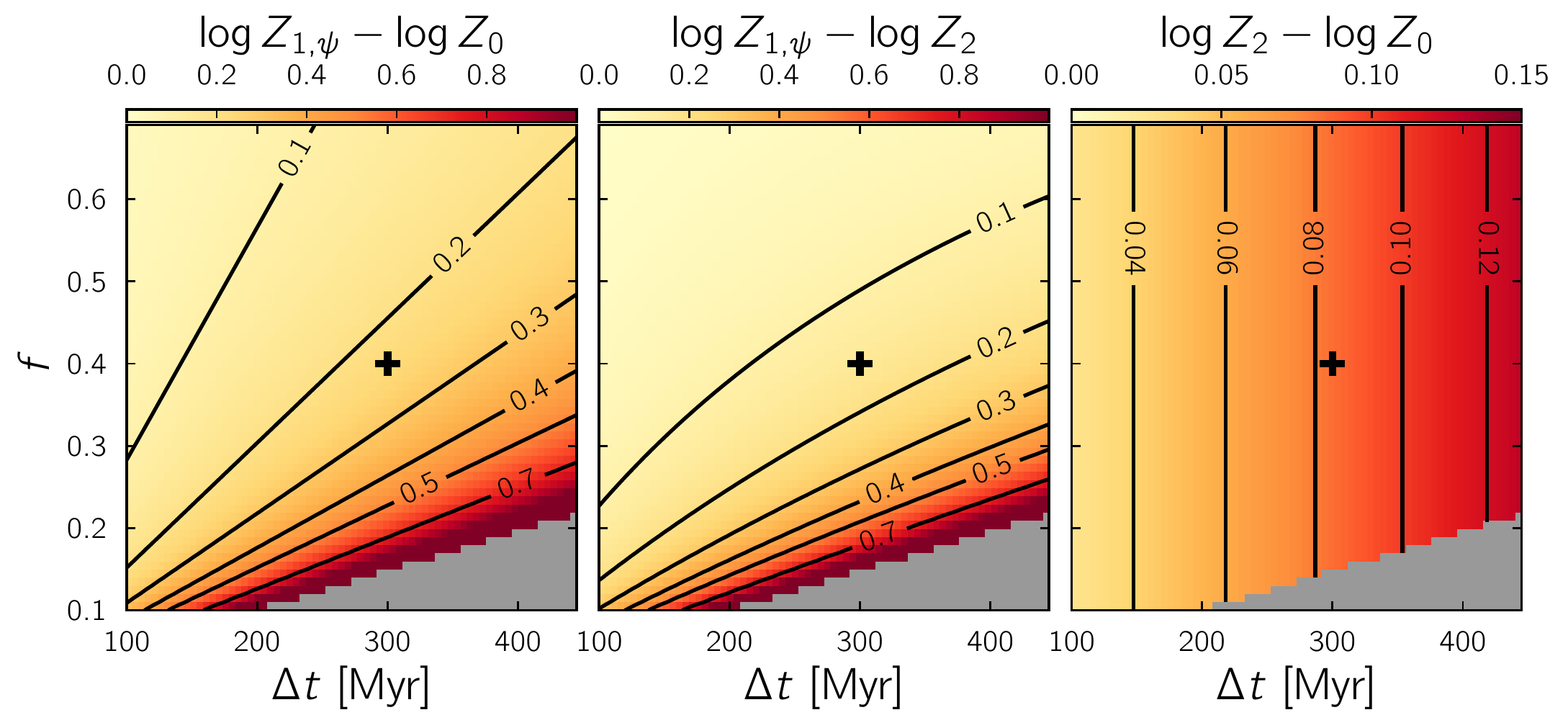}
\caption{This three panels explore the simple chemical evolution model described in Section~5.2. The panels show how the model predictions change with changing $f$ and $\Delta t$. The black crosses mark the fiducial values for $f$ and $\Delta t$. The other two parameters, $y_o$ and $R$, are fixed at their fiducial values.}\label{figb1}
\end{figure*}

\begin{figure*}
\centering
\includegraphics[width=0.7\textwidth]{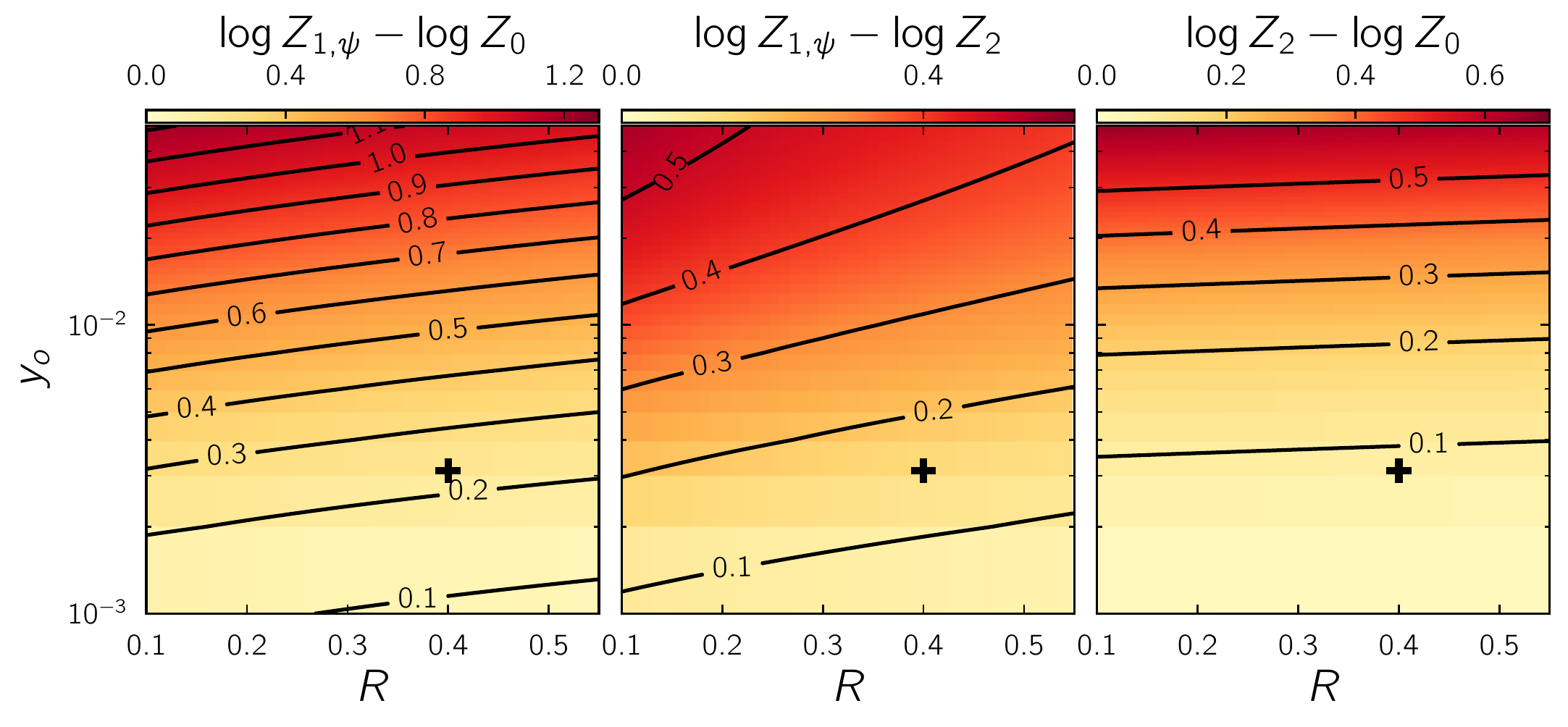}
\caption{Same as Figure~\ref{figb2} but with different $y_o$ and $R$. The black crosses mark the fiducial values for $y_o$ and $R$. The other two parameters, $f$ and $\Delta t$, are fixed at their fiducial values. }\label{figb2}
\end{figure*}

In Section~5.2, we calculate the chemical evolution model using a set of fiducial values for the model parameters, i.e.~$(\Delta t, f, R, y_o) = (300\ {\rm Myr}, 0.4, 0.4, 0.00313)$. In Figures~\ref{figb1} and \ref{figb2}, we explore how our model behaves around the fiducial values. We note that in our model, $\Delta t$ degenerates with $\psi$, i.e. doubling the star formation rate has the same effect as doubling the dynamical time because the same amount of oxygen is produced. 

In Figure~\ref{figb1}, we fix $R$ and $y_o$ to their fiducial values and vary $\Delta t$ and $f$. At fixed $\log Z_{1,\phi} - \log Z_0$ and $\log Z_{1,\phi} - \log Z_2$ values (left and middle panels), there are strong positive correlations between $\Delta t$ and $f$. This is because increasing $\Delta t$ and thus the amount of oxygen produced can be compensated by distributing the oxygen into a larger box, i.e.~a larger volume with mass fraction $f$. Over the plausible dynamical time of NGC\,1365 (200 to 400~Myr; Section~5.1), $f$ of approximately 0.3 to 0.6 is required to reproduce our observation. This range can be compared with the fact that about 10 to 30\% of the disk area in typical star-forming galaxies is occupied by \HII\ regions at any instance. It is also possible that $f$ intrinsically correlates with $\Delta t$ because a larger $\Delta t$ would allow more time for mixing to chemically homogenize more gas. In the right panel of Figure~\ref{figb1}, $\log Z_2 - \log Z_0$ does not depend on $f$. This is because by $t_2$ the oxygen produced inside the volume effected by mixing is all returned to the spatial element considered. 

In Figure~\ref{figb2}, we fix $\Delta t$ and $f$, and vary $R$ and $y_o$. The resulting abundance differences are relatively insensitive to the return mass fraction $R$ but depends strongly on the oxygen yield $y_o$. In this work, we use the empirically constrained yield from \citet{Kudritzki:2015fp}. Theoretical oxygen yields are considerably higher than the value we adopted. \citet{Vincenzo:2016hb} compile oxygen yields from different stellar models and show that the oxygen yields differ significantly from model to model (see \citealt{Vincenzo:2016hb} and references therein). The oxygen yield is sensitive to the assumed initial mass function and upper mass cut-off. The theoretical oxygen yields span a wide range from 0.007 to 0.04. In Figure~\ref{figb2}, we show that if the yield is 0.04 then one expects approximately 1~dex oxygen abundance difference when gas goes through the inter-arm region (left panel). Even after mixing, one expects a 0.5~dex increase in oxygen abundance only after the gas orbits a half of the galaxy (right panel). Such a large change in the oxygen abundances is not observed. 

When deriving equations~\ref{eq-deltaz10-simple}, \ref{eq-deltaz12-simple} and \ref{eq-deltaz20-simple}, we assume that $Z(t)$ in the third right-hand-side term of equation~2 can be approximated by a constant, $Z(t)\approx m_{o,0}/m_{g,0}$, which we adopt the mean observational value from \citet{Bresolin:2005uq}. However, $Z(t)$ is both a function of azimuthal angle (and thus time) and radial distance. If we repeat the calculation with 0.15~dex higher and lower $Z(t)$, we reach ($\log{Z_{1,\psi}} - \log{Z_0}$, $\log{Z_{1,\psi}} - \log{Z_2}$, $\log{Z_2} - \log{Z_0}$) of (0.17,0.11,0.06) and (0.27,0.17,0.10), respectively. Compared to (0.23,0.15,0.08) in equations~16, 17 and 18, the differences are approximately 0.05~dex. The constant approximation of $Z(t)$ is not unreasonable given that the measurement errors are about 0.3~dex for individual \HII\ regions. 

Finally, a direct prediction of our model is that the amplitude of the oxygen abundance variations should depend on the inter-arm traveling time. Assuming that the spiral arm pattern speed is constant, the increase of inter-arm traveling time with radius should yield larger abundance variations at larger radii. However, this effect can be partly compensated by the lower star formation rate (less oxygen produced) and lower oxygen abundance (see above) at larger radii. Investigating how the abundance variation correlates with radius using a more sophisticated chemical evolution model is beyond the scope of this work. Oxygen abundances measured at a higher accuracy will also be needed to know whether the amplitude of the variations changes with radius.

\end{CJK*}
\end{document}